\documentclass[%
 reprint,
superscriptaddress,
showpacs,
 amsmath,amssymb,
 aps,
floatfix,
]{revtex4-1}
\usepackage{graphicx}
\usepackage{dcolumn}
\usepackage{bm}
\usepackage[hidelinks]{hyperref}
\hypersetup{
	colorlinks=true,
	linkcolor=red,
	citecolor =red,      
	urlcolor=blue
}
\usepackage{xcolor}

\begin{document}
\title{Power loss of hot Dirac fermions in silicene and its near equivalence with graphene}
\author{S.~S.~Kubakaddi}
\thanks{Corresponding author}\email{sskubakaddi@gmail.com}
\affiliation{Department of Physics, K. L. E. Technological University, Hubballi-580~031, Karnataka, India}
\author{Huynh~V.~Phuc}
\affiliation{Division of Theoretical Physics, Dong Thap University, Cao~Lanh~870000, Vietnam}

\date{\today}

\begin{abstract}
The power loss $P$ of hot Dirac fermions through the coupling to the intrinsic intravalley and intervalley acoustic and optical phonons is analytically investigated in silicene as a function of electron temperature $T_e$ and density $n_s$.  At very low $T_e$, the power dissipation is found to follow the Bloch-Gr\"{u}neisen power-law $\propto T_e^4$ and $n_s^{-0.5}$, as in graphene, and for $T_e \lesssim20-30$~K, the power loss is predominantly due to the intravalley acoustic phonon scattering. On the other hand, dispersionless low energy intervalley acoustic phonons begin to dominate the power transfer at temperatures as low as $\sim$$30$~K, and optical phonons dominate at $T_e \gtrsim200$~K, unlike the graphene. The total power loss increases with $T_e$ with a value of $\sim$$10^{10}$~eV/s at $300$~K, which is the same order of magnitude as in graphene. The power loss due to intravalley acoustic phonons increases with $n_s$ at higher $T_e$, whereas due to the intervalley acoustic and optical phonons is found to be independent of $n_s$. Interestingly, the energy relaxation time in silicene is about $4$ times higher than that in graphene. For this reason, silicene may be superior over graphene for its applications in bolometers and calorimeters. Power transfer to the surface optical phonons $P_{\text{SO}}$ is also studied as a function of $T_e$ and $n_s$ for silicene on Al$_2$O$_3$ substrate and it is found to be greater than the intrinsic phonon contribution at higher $T_e$. Substrate engineering is discussed to reduce $P_{\text{SO}}$.

\end{abstract}

\maketitle


\section{Introduction}

Silicene, a monolayer of silicon atoms arranged in honeycomb lattice in two-dimensions (2D), has generated a strong interest in the condensed matter physics community~\cite{prl102(2009)236804,apl97(2010)223109,prl107(2011)076802,prb85(2012)075423,nl12(2012)3507,prl108(2012)155501,apl102(2013)043113,apl103(2013)261904,prb87(2013)115418,nnano10(2015)227,nnano10(2015)202,sst31(2016)065012,sst31(2016)115004,prb93(2016)035414,prb93(2016)155413,sst33(2018)065011,prb94(2016)075409,pms83(2016)24,jap124(2018)044306,jms1199(2020)126878,prb101(2020)205408}. Because of the similarity of the lattice structures, the band structure of silicene is similar to that of graphene, and charge carriers are massless relativistic Dirac fermions.  This material has drawn much attention because of its hopeful applications in Si nanoelectronics due to its compatibility with the existing silicon-based electronics technology and a tunable band-gap that can be created at room temperature by application of an electric field perpendicular to the monolayer of the atoms~\cite{prb85(2012)075423}. Moreover, because of the larger spin-orbit coupling, the spin quantum Hall effect is shown to be stronger~\cite{prl107(2011)076802,apl102(2013)043113}.

Even though the possibility of obtaining free-standing silicene has been demonstrated theoretically~\cite{apl103(2013)261904}, it has not been synthesized yet. However, silicene is grown on the surface of Ag(111)~\cite{apl97(2010)223109,nl12(2012)3507,prl108(2012)155501}. The first silicene field-effect transistor (FET) on a polar substrate Al$_2$O$_3$, operating at room temperature, was successfully demonstrated by Tao {\it et al.}~\cite{nnano10(2015)227} with an on/off ratio $\sim$$10$, and hinting at the existence of a bandgap of about $\sim$$210$~meV~\cite{nnano10(2015)227,nnano10(2015)202}.

Characterization of electronic transport of the material is critical for assessing and understanding its potential significance for its applications in electronic devices. In suspended silicene, phonon limited electron mobility and velocity-field characteristics are the most theoretically studied electronic transport properties by considering the electron scattering by intrinsic intravalley and intervalley acoustic phonons (ap) and optical phonons (op)~\cite{prb87(2013)115418,sst31(2016)065012,sst31(2016)115004,prb93(2016)035414,sst33(2018)065011}. The phonon branches taken into account are the longitudinal acoustic (LA), transverse acoustic (TA), out-of-plane acoustic (ZA), longitudinal optical (LO), transverse optical (TO) and out-of-plane optical (ZO). Employing the full-band Monte Carlo (FMC) model~\cite{prb87(2013)115418} and the analytical Monte Carlo (AMC) model~\cite{sst31(2016)065012,sst31(2016)115004,sst33(2018)065011}, the predicted room temperature mobility is $\sim$$1000$~cm$^2$/Vs, and the saturation drift velocity is about $5-10\times 10^6$~cm/s.  From the first principle calculations, using the FMC model Li {\it et al.}~\cite{prb87(2013)115418} have determined the phonon energies and deformation potential coupling constants for the intravalley and intervalley LA, TA, and ZA phonons and intravalley and intervalley LO, TO, and ZO phonons. However, the experimental work on these transport properties is scarce, with the first FET on Al$_2$O$_3$ exhibiting a room temperature mobility of $\sim$$100$~cm$^2$/Vs~\cite{nnano10(2015)227}. Scattering by surface polar optical (SO) phonons is considered in an attempt to obtain this value~\cite{sst31(2016)065012}.

Because of its buckled structure, in suspended silicene the electron scattering by out-of-plane acoustic flexural (ZA) modes, due to their parabolic dispersion, is very strong~\cite{prb93(2016)155413} giving extremely low mobility $\sim$$10^{-3}$~cm$^2$/Vs at $300$~K. Different damping models, yet elusive, of ZA phonon scattering, have been rigorously discussed~\cite{prb93(2016)155413,sst33(2018)065011,jap124(2018)044306}. Better mobilities \mbox{$701$~cm$^2$/Vs}~\cite{jap124(2018)044306} and $1200$~cm$^2$/Vs~\cite{sst33(2018)065011} and saturation velocity ($\sim$$8\times 10^6$~cm/s)~\cite{sst33(2018)065011}, useful for practical purposes, are obtained. For applications in electronic devices, there is a need for suppressing the scattering by out-of-plane ZA phonons with parabolic dispersion in buckled silicene. It is to be noted that in the theoretical work of Refs.~\cite{prb87(2013)115418,sst31(2016)065012,sst31(2016)115004,sst33(2018)065011}, the ZA phonons were artificially ‘regularized’ by approximating the parabolic dispersion with an ad hoc linear dispersion at long wavelengths. This model is accepted in our present work.

Amongst the transport properties, what remains to be investigated theoretically is the power dissipation of hot Dirac fermions to the lattice in silicene. In photoexcited samples and the samples subject to high electric fields, due to the electron-electron interactions occurring at much faster time scale than the electron-phonon processes, the electron system establishes its internal thermal equilibrium at an electron temperature $T_e$ greater than the lattice temperature $T$ and electrons are driven out of equilibrium with the lattice. In steady-state, these electrons will relax towards equilibrium with lattice by losing energy with phonons as the cooling channels. This is an important phenomenon as it affects thermal dissipation and heat management which are key issues in nanoscale electronic devices. A quantitative understanding of this hot electron (Dirac fermions) cooling power (i.e. power loss) in silicene is of central importance as the device performance in the high field is determined by these hot electrons.  In addition, this property is crucial for applications in a variety of devices such as calorimeters, bolometers, infrared and THz detectors. 

In this work, we analytically study the hot Dirac fermion power loss due to intrinsic acoustic and optical phonons in suspended silicene and due to the surface polar optical phonons in samples on the substrate. We present a general theory for the electron cooling power via the different phonon baths, including surface phonons, in Section 2.  The results and discussion of our numerical calculations of the cooling powers as a function of electron temperature $T_e$ and electron density $n_s$ are presented in Section 3.  Conclusions are given in Section 4.
\section{Analytical model for the power loss}
The energy band structure of the silicene is given by the Dirac cone analytical relation $E_\text{k}=\hbar v_F|\bf{k}|$ with the density states $D(E_\text{k})=gE_\text{k}/[2\pi(\hbar v_F)^2]$, where $E_\text{k}$ is the electron energy, $\bf{k}$ is the wave vector, $g=g_vg_s$, $g_v(g_s)$ is the valley (spin) degeneracy, and $v_F=5.8\times 10^7$~cm/s~\cite{prb87(2013)115418} is the Fermi velocity in silicene. This approximation of linear dispersion is limited for the electron energy up to $0.2$~eV~\cite{sst31(2016)065012,sst33(2018)065011} and the range of electric field considered in the present work is such that the average electron energies are below this value. The electron distribution  is assumed to be  given by the heated  Fermi-Dirac distribution $f(E_\text{k})=\{{\rm exp}[(E_{\text{k}}-\mu)/k_BT_e]+1\}^{-1}$, where $\mu$ is the chemical potential  of the Dirac fermions determined by 2D electron density $n_s=\int f(E_\text{k})D(E_\text{k})dE_{\text{k}}$. We investigate the electron cooling power $P$ in suspended silicene considering the electron interaction with the intrinsic acoustic and optical phonons as the cooling channels. Electrons are assumed to couple with phonons of energy $\hbar\omega_{q\lambda}$ with branch index $\lambda$ and wave vector ${\bf q}$ via acoustic and optical intravalley and intervalley deformation potential coupling~\cite{prb87(2013)115418}. Electron interactions with phonons close to the $\Gamma$ point of the Brillouin zone can be considered as intravalley ($\Gamma$ point wave vector)  scattering, while interacting with phonons close to $K$ point can be regarded as intervalley ($K$ point wave vector) scattering causing $K\to K'$ transition events. In our model, we include intrinsic intravalley and intervalley acoustic phonons ($\lambda=$ LA, TA, and ZA) and intravalley and intervalley optical phonons ($\lambda=$ LO, TO, and ZO). Linear dispersion $\omega_{q\lambda} = v_{\lambda}q$, where $v_{\lambda}$ is the acoustic phonon velocity of the branch $\lambda$, is assumed for the intravalley acoustic phonon scattering, including ZA phonons ‘regularized’~\cite{prb87(2013)115418,sst31(2016)065012,sst31(2016)115004,sst33(2018)065011}. While, for intervalley LA, TA and ZA modes and for intravalley and intervalley optical modes $\hbar\omega_{q\lambda}$ is dispersionless and take the respective constant phonon energy values $\hbar\omega_{q\lambda}=\hbar\omega_{0\lambda}$~\cite{prb87(2013)115418}. Moreover, we also present $P$ calculations for silicene on the Al$_2$O$_3$ substrate, in which case scattering by surface polar optical (SO) phonons of energy $\hbar\omega_{\text{SO}}$ provides an additional cooling bath.

The hot electron power loss  to the phonons of branch $\lambda$ is given by $P_{\lambda}=(1/N_e)\sum_{\text{q}}\hbar\omega_{\text{q}\lambda}(dN_{\text{q}\lambda}/dt)_{\text{el-ph}}$, where $N_e$ is the total number of electrons and $(dN_{\text{q}\lambda}/dt)_{\text{el-ph}}$ is the rate of change of phonon distribution $N_{\text{q}\lambda}$ due to electron-phonon (el-ph) coupling. The phonon rate equation  $(dN_{\text{q}\lambda}/dt)_{\text{el-ph}}$, using Fermi golden rule, is given by 
\begin{align}
\left(\frac{dN_{\text{q}\lambda}}{dt}\right)_{\text{el-ph}}&=\frac{2\pi g}{\hbar}\sum_{\bf k}|M_{\lambda}(q)|^2F_{\lambda}(\text{k},\text{k'})\nonumber\\
&\quad\times\delta(E_{\text{k'}}-E_\text{k}-\hbar\omega_{\text{q}\lambda}),
\end{align}
where a composite Fermi-boson distribution is defined as $F_{\lambda}(\text{k},\text{k'})=\{(N_{\text{q}\lambda}+1)f(E_\text{k}+\hbar\omega_{\text{q}\lambda})[1-f(E_\text{k})]-N_{\text{q}\lambda}f(E_\text{k})[1-f(E_\text{k}+\hbar\omega_{\text{q}\lambda})\}$, which is zero when $T_e=T$.
\subsection{Electron power loss due to intravalley acoustic phonons }
The electron cooling power due to the intravalley ($\Gamma$ point) acoustic phonon ($\lambda=$ LA, TA, and ZA) scattering is obtained using the electron-acoustic phonon matrix element $|M_{\lambda}(q)|^2=|g_{\lambda}(q)|^2G(\theta)$, where $|g_{\lambda}(q)|^2=(D_{1\lambda}^2\hbar q/2A\rho_m v_{\lambda})$~\cite{prb87(2013)115418,prb77(2008)115449}, $D_{1\lambda}$ is the first-order acoustic deformation potential coupling constant, $\rho_m$ is the areal mass density of the silicene, $A$ is the area, $\theta$ is the angle between initial ${\bf k}$ and final ${\bf k'}$ wave vectors of the electron and we define the chiral function $G(\theta)=(1+\cos\theta)/2$ arising from the spinor wave function of the electron. The power loss $P_{\text{ap},\lambda}$ due to intravalley acoustic phonons is given by~\cite{prb79(2009)075417}
\begin{align}
P_{\text{ap},\lambda}&=-\frac{gD_{1\lambda}^2}{4\pi^2n_s\rho_m\hbar^5v_{\lambda}^3v_F^3}\int_0^{\infty}d(\hbar\omega_{\text{q}\lambda})(\hbar\omega_{\text{q}\lambda})^2\nonumber\\
&\quad\times\int_{\gamma}^{\infty}dE_{\text{k}}\frac{(E_{\lambda}+\hbar\omega_{\text{q}\lambda})G(E_q,E_k)}{[1-(\gamma/E_k)^2]^{1/2}} [N_{q\lambda}(T_e)\nonumber\\
&\quad-N_{q\lambda}(T)][f(E_k)-f(E_k+\hbar\omega_{q\lambda})],
\end{align}
where $\gamma=E_q/2$ with $E_q=\hbar v_Fq$, $N_{q\lambda}=[{\rm exp}(\hbar\omega_{q\lambda}/k_BT)-1]^{-1}$ is the Bose-Einstein distribution at lattice temperature $T$, and $G(\theta)=G(E_k,E_q)=[1-(\gamma/E_k)^2]$ in the quasi-elastic approximation. Here, the screening of el-ph interaction is ignored as it is theoretically and experimentally justified for deformation potential coupling in graphene~\cite{prb77(2008)115449,prb79(2009)075417,prb79(2009)235406,prb81(2010)245404,prl105(2010)256805,prl109(2012)056805,prb85(2012)115403}.

At very low-temperature $T$, $T_e\ll T_{\text{BG}}$, where $T_\text{BG}=(2\hbar v_{\lambda}k_F/k_B)$ is the Bloch-Gr\"{u}neisen (BG) temperature, the cooling power is given by the simple power law~\cite{prb79(2009)075417} 
\begin{equation}\label{eq2}
P_{\text{ap},\lambda}=P_{0\lambda}(T_e^4-T^4)/n_s^{1/2},
\end{equation}
with $P_{0\lambda}=P_0(D_{1\lambda}^2/v_\lambda^3)$ and $P_0=(\pi^{5/2}k_B^4)/(15\rho_m\hbar^4v_F^2)$.
\subsection{Electron power loss due to intervalley acoustic and intravalley and intervalley optical phonons (dispersionless phonons)}
The theory of electron power loss due to optical phonons via optical deformation potential coupling in graphene is given in Refs.~\cite{prb79(2009)235406,prb81(2010)245404,prb86(2012)045413}. We have obtained, following Refs.~\cite{jap113(2013)063705,jpcm30(2018)265303} an expression for the electron cooling power $P_{\text{ap,op},\lambda}$ due to intervalley ($K$ point) acoustic phonons (LA, TA, and ZA) and intravalley ($\Gamma$ point)  and intervalley ($K$ point)  optical phonons (LO, TO, and ZO) taking account of ‘hot phonon effect’. Employing the matrix element $|g_{\lambda}(q)|^2=(D_{0\lambda}^2\hbar/2A\rho_m\omega_{0\lambda})$~\cite{prb87(2013)115418}, where $\hbar\omega_{q\lambda}=\hbar\omega_{0\lambda}$ is the phonon energy of $\lambda$th mode, $D_{0\lambda}$ is the zeroth order deformation potential constant, the power loss $P_{ap,op,\lambda}$ is given by
\begin{align}\label{eq3}
P_{\text{ap,op},\lambda}&=-\frac{\hbar\omega_{0\lambda}}{2\pi n_s(\hbar v_F)^2}\int_{E_{ql,\lambda}}^{\infty}dE_qE_q[(N_{q\lambda}+1)e^{-\beta\hbar\omega_{0\lambda}}\nonumber\\
&\quad-N_{q\lambda}]\Gamma_{\lambda}(q),
\end{align}
where $\beta=1/(k_BT_e)$, 
\begin{align}
\Gamma_{\lambda}(q)&=\frac{gD_{0\lambda}^2\hbar}{2\pi\rho_m(\hbar v_F)^2(\hbar\omega_{0\lambda})}I_{\lambda}(E_q,\hbar\omega_{0\lambda}),\\
\label{eq4a}I_{\lambda}(E_q,\hbar\omega_{0\lambda})&=\int_{E_{kl\lambda}}^{\infty}dE_kE_kG_{\lambda}(E_k,E_q)Z_{\lambda}(E_k,E_q)\nonumber\\
&\quad\times\frac{(E_k+\hbar\omega_{0\lambda})}{E_kE_q}f(E)k)[1-f(E_k+\hbar\omega_{0\lambda})],\\
Z_{\lambda}(E_k,E_q)&=\frac{2E_kE_q}{\sqrt{4E_k^2E_q^2-[(\hbar\omega_{0\lambda})^2+2E_k\hbar\omega_{0\lambda}-E_q^2]^2}},\\
E_{kl\lambda}&=(E_{\text{q}}-\hbar\omega_{0\lambda})/2,\text{ and }  E_{ql\lambda}=\hbar\omega_{0\lambda}.
\end{align}
For optical phonons in graphene $G_{\lambda}(E_k,E_q)=(1/2)\{1\pm[(E_k^2-E_q^2)+(E_k+\hbar\omega_{0\lambda})^2]/2E_k(E_k+\hbar\omega_{0\lambda})\}$~\cite{prb86(2012)045413,prb81(2010)195442}. However, we have set $|G_{\lambda}(E_k,E_q)|=1$ in our calculations, matching with Li {\it et al.}~\cite{prb87(2013)115418}, as its form is clearly not known for all the dispersionless phonons considered here.  The ‘hot phonon’ distribution function $N_{q\lambda}$, obtained by solving the phonon Boltzmann transport equation in the relaxation time approximation, is given by $N_{\text{q}\lambda}=[N_{\text{q}\lambda}^0+\tau_{p\lambda}\Gamma_{\lambda}(q){\rm exp}(-\beta\hbar\omega_{0\lambda})]/\{1+\tau_{p\lambda}\Gamma_{\lambda}(q)[1-{\rm exp}(-\beta\hbar\omega_{0\lambda})]\}$, where $N_{\text{q}\lambda}^0$ is the Bose-Einstein distribution function for phonons at lattice temperature $T$  and $\tau_{p\lambda}$ is the relaxation time of the $\lambda$th branch phonon.
\subsection{Electron power loss due to surface polar  optical (SO) phonons}
For a silicene on substrate Al$_2$O$_3$~\cite{nnano10(2015)227,prb94(2016)075409}  it has been shown that SO phonon scattering contributes significantly and degrades electron mobility~\cite{sst31(2016)065012}. The power loss due to SO phonons, $P_{\text{SO}}$, has been studied for monolayer  graphene on substrate, without considering the hot phonon effect~\cite{prb86(2012)045413,prl104(2010)236601},  and for  bilayer graphene with hot phonon effect~\cite{jap113(2013)063705}. We obtain an expression for $P_{\text{SO}}$, taking account of screening and the hot phonon effect. In graphene, the electron-SO (el-so) phonon interaction  matrix element for the SO phonons of energy $\hbar\omega_{\text{SO}}$ is given by  $|g_{\text{SO}}(q)|^2=e^2F^2[{\rm exp}(-2qd)]/q$~\cite{prb82(2010)115452}, where $F^2=(2\pi\hbar\omega_{\text{SO}}/A)\epsilon_p$ is the square of the Fr\"{o}lich coupling constant for SO phonon, $\epsilon_p=[(\epsilon_{ox}^{\infty}+1)^{-1}-(\epsilon_{ox}^0+1)^{-1}]$, $\epsilon_{ox}^0(\epsilon_{ox}^{\infty})$ is the static (high frequency) dielectric constant and $d$  is the distance of the substrate from silicene layer. The expression for $P_\text{SO}$ is found to be the  same as Eq.~\eqref{eq3}, with replacement of  $\hbar\omega_{0\lambda}$ by $\hbar\omega_{\text{SO}}$ and $\Gamma_{\lambda}$ by 
\begin{align}
\Gamma_{\text{SO}}(q)&=\frac{2ge^2\hbar\omega_{\text{SO}}\epsilon_p}{\hbar^2 v_F}\left(\frac{e^{-2E_q(d/\hbar v_F)}}{E_q\epsilon^2(q)}\right)I(E_q,\hbar\omega_{\text{SO}}).
\end{align}
Here $I(E_q,\hbar\omega_{\text{SO}})$ is the same as Eq.~\eqref{eq4a} with $\hbar\omega_{0\lambda}$ replaced by $\hbar\omega_{\text{SO}}$ and $|G_{\text{SO}}(E_k,E_q)|=(1/2)\{1+[(E_k^2-E_q^2)+(E_k+\hbar\omega_{\text{SO}})^2]/2E_k(E_k+\hbar\omega_{\text{SO}})\}$. The temperature dependent static screening function is given by $\epsilon(q,T)=[1+(q_s/q)]$~\cite{prb86(2012)045413,prb82(2010)115452}, where the screening wave vector $q_s=(2\pi e^2/\epsilon_s)\int (\partial f/\partial E)D(E)dE$~\cite{prb86(2012)045413} with $\epsilon_s=(\epsilon_{ox}^0+1)/2$, the dielectric constant average of the substrate and vacuum~\cite{prb82(2010)115452}. It is to be noted that there are two surface phonons SO$1$ and SO$2$ of different energy for substrate Al$_2$O$_3$~\cite{prb82(2010)115452}.
\section{Results and discussion}
In this section we present the numerical calculations of hot  electron power loss  due to different competing channels in silicene for the electron temperature range $T_e =1-300$~K and electron density $n_s = 1-10$~$n_0$, where $n_0 =1\times10^{12}$/cm$^2$ throughout the discussion.  The power loss  is presented at two  lattice   temperatures $T= 0.1$~K and $4.2$~K. The material parameters of the silicene used are~\cite{prb87(2013)115418}: $\rho_m=7.2\times 10^{-8}$~g/cm$^2$, $g=4$, $v_{\text{LA}}=8.8\times 10^5$~cm/s, $v_{\text{TA}}=5.4\times 10^5$~cm/s, $v_{\text{ZA}}=0.63\times 10^5$~cm/s, and $v_F=5.8\times 10^7$~cm/s for the suspended sample. The deformation potential constants $D_{1\lambda}$ and $D_{0\lambda}$ and phonon energies $\hbar\omega_{0\lambda}$ taken from Li {\it et al.}~\cite{prb87(2013)115418} are listed in Table~\ref{table1}. For the  sample supported on Al$_2$O$_3$ substrate, the two surface polar optical phonon   energies ($\hbar\omega_{\text{SO}1}$ and $\hbar\omega_{\text{SO}2}$) and  the dielectric constants of the substrate given in Table~\ref{table2} are taken from Refs.~\cite{prb82(2010)115452,jap90(2001)4587}, with the modified Fermi $v_F=3.4\times 10^7$~cm/s~\cite{sst31(2016)065012}. While considering the hot phonon effect, $\tau_p = 0,1$ and $5$~ps are chosen for illustration for all the optical phonons and intervalley acoustic phonons. We note that the phonon relaxation times $\tau_p$ are found to be  different for different branches  and are  in the range $1-1.5$~ps except for  LA and   ZO phonon modes at $K$ point for which  $\tau_p = 2.5$ and $5.0$~ps, respectively~\cite{jpdap51(2018)415102}.

\begin{table}[!b]
	\caption{\label{table1}%
		Phonon energies at the $\Gamma$ and $K$ symmetry points  and extracted deformation potential constants for electron-phonon interaction for monolayer silicene (Li {\it et al.}~\cite{prb87(2013)115418}).
	}
	\begin{ruledtabular}
		\begin{tabular}{||l|d|d|l|l||}
			\multicolumn{1}{||c|}{Phonon}&\multicolumn{2}{c|}{Phonon energy}&\multicolumn{2}{c||}{Deformation potential}\\
			\multicolumn{1}{||c|}{modes}&\multicolumn{2}{c|}{(meV)}&\multicolumn{2}{c||}{constants}\\
			\colrule
			&	\multicolumn{1}{c|}{}&	\multicolumn{1}{c|}{}&	\multicolumn{1}{c|}{Intravalley}&	\multicolumn{1}{c||}{Intervalley}\\
				&\multicolumn{1}{c|}{$\Gamma$}&	\multicolumn{1}{c|}{$K$}&	\multicolumn{1}{c|}{$D_1$, $D_0$}&	\multicolumn{1}{c||}{$D_0$(eV/cm)}\\
			\colrule
			LA&0&13.2&3.2~eV&$4.2\times 10^7$\\
			TA&0&23.7&8.7~eV&$1.4\times 10^8$\\
			ZA&0&13.2&2.0~eV&$6.1\times 10^7$\\
			LO&68.8&61.7&$1.9\times 10^8$~eV/cm&$1.7\times 10^8$\\
			TO&68.8&50.6&$1.8\times 10^8$~eV/cm&$1.4\times 10^8$\\
				ZO&22.7&50.6&$6.3\times 10^7$~eV/cm&$4.3\times 10^7$\\
		\end{tabular}
	\end{ruledtabular}
\end{table}

\begin{table}[!b]
	\caption{\label{table2}%
		Material parameters of Al$_2$O$_3$ substrate~\cite{prb82(2010)115452,jap90(2001)4587}.  
	}
	\begin{ruledtabular}
		\begin{tabular}{||l|l||}
			\multicolumn{1}{||c|}{Quantities}&\multicolumn{1}{l||}{Values}\\
			\colrule
		Fermi velocity $v_F$&$3.4\times 10^7$~cm/s\\
		Average dielectric constant $\epsilon_s$&6.76\\
		Low frequency dielectric constant $\epsilon_{0x}^0$&12.53\\
		High frequency dielectric constant $\epsilon_{0x}^{\infty}$&3.2\\
		Surface optical phonon energy $\hbar\omega_{\text{SO}1}$&55.01~meV\\
		Surface optical phonon energy $\hbar\omega_{\text{SO}2}$&94.29~meV\\
		\end{tabular}
	\end{ruledtabular}
\end{table}
\subsection{Power loss due to intrinsic intravalley and intervalley acoustic and optical phonons}
First we explore the temperature dependence of the power loss  $P_{\text{ap}}$ due to intravalley acoustic phonon scattering in the range $T_e =1-100$~K at $T = 0.1$~K.  At low temperatures, where the thermal energy of the electron distribution is much smaller than the intervalley acoustic and intravalley and intervalley optical phonon energies, the heat dissipation is dominated by the intravalley acoustic phonon scattering. In this regime, the cooling power due to each of the intravalley acoustic phonons scattering can be described by the generic power-law behavior  $P_{{\rm ap},\lambda}=\Sigma_{\lambda}(\mu,T_e)(T_e^{\delta}-T^{\delta})$~\cite{prb79(2009)075417,prb81(2010)245404}, where $\Sigma_{\lambda}$ is the coupling constant that depends on the chemical potential $\mu$ and the electron temperature $T_e$, and $\delta$ is the exponent of the power-law which overall decreases with increasing temperature. These dependencies of $\Sigma_{\lambda}$ and $\delta$ are determined by the el-ph matrix element and the composite Fermi-boson distribution function $F_{\lambda}(\bf {k}, \bf{k'})$.

At very low temperature $T$, $T_e\ll T_{\text{BG}}$, from  Eq.~\eqref{eq2}, $P_{ap}\propto T_e^4$ and $n_s^{-1/2}$. Expressing $n_s$ in  terms of $n_{s0}$, where $n_s= n_{s0}\times10^{12}$~cm$^{-2}$, we find $P_{0\text{LA}} = 5.46\times10^{-18}/n_{s0}^{1/2}$, $P_{0\text{TA}}=1.75\times10^{-16}/n_{s0}^{1/2}$, and $P_{0\text{ZA}}= 5.81\times10^{-15}/n_{s0}^{1/2}$~W/K$^{4}$-cm.   For graphene, for the parameters in Ref.~\cite{prb79(2009)075417}, $P_{0\text{LA}} = 5.23\times 10^{-18}/n_{s0}^{1/2}$~W/K$^4$-cm, which is nearly same as  the $P_{0\text{LA}}$ in silicene. In silicene $T_{\text{BG}}=23.83\sqrt{n_{s0}}$, $14.62\sqrt{n_{s0}}$, and $1.706\sqrt{n_{s0}}$, respectively, for LA, TA and ZA phonons, where as in graphene $T_{\text{BG}} = 54.15\sqrt{n_{s0}}$ for LA phonons~\cite{prb79(2009)075417}.
 
\begin{figure}[!t]
	\includegraphics[width=\columnwidth]{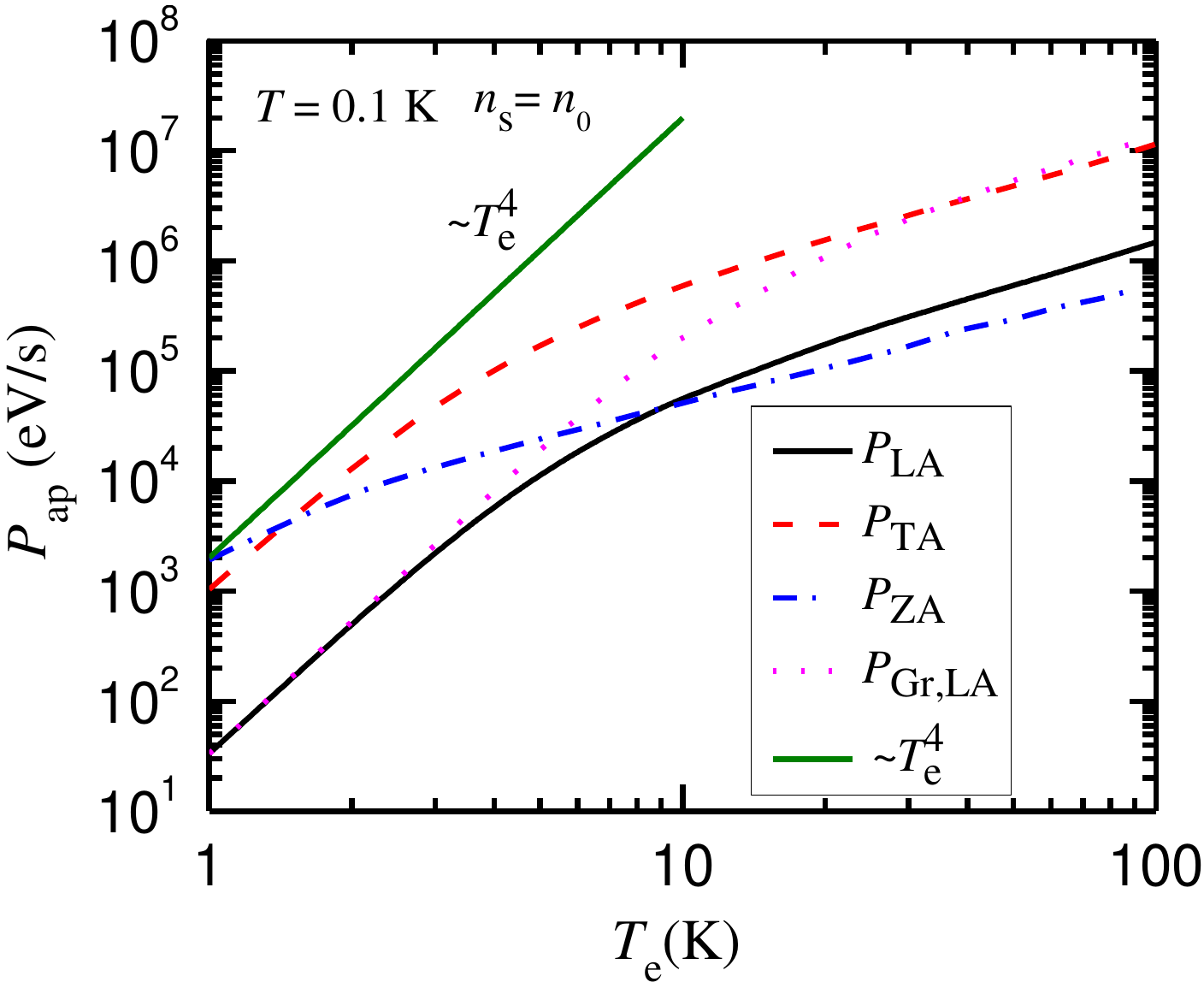}
	\caption{Power loss $P_{\text{ap}}$ due to intravalley acoustic phonons as a function of electron temperature  $T_e$, at lattice temperature $T = 0.1$~K. Curve $P_{\text{Gr, LA}}$  is the power loss in graphene.}
	\label{fig:Fig1}
\end{figure}

In Fig.~\ref{fig:Fig1} the power loss $P_{\text{ap}}$ from the intravalley acoustic phonon scattering, with chiral function, is depicted as a function of electron temperature $T_e$ for $n_s = n_0$ at $T=0.1$~K. $P_{ap}$ curves due to LA, TA, and ZA phonons are shown along with the curve for $T_e^4$ power law. For comparison $P_{\text{ap}}$ due to LA phonons in graphene from Ref.~\cite{prb79(2009)075417} is also shown. The $T_e^4$ power law is obeyed by LA (TA) phonons for about $T_e < 3~(2)$~K and ZA phonons do not seem to show this power law in the $T_e$ region shown, as its $T_{\text{BG}}$ is too small. It is found that for about $T_e < 2$~K, $P_{\text{ZA}}$ is dominant, although coupling constant is smaller, which can be attributed to the small phonon energy (i.e. a large phonon occupation number) near the zone center because of the small $v_{\text{ZA}} = 0.63\times10^5$~cm/s (about an order of magnitude smaller than $v_{\text{LA}}$ and $v_{\text{TA}}$). While for $T_e \gtrsim2$~K, with increasing temperature,  $P_{\text{TA}}$ is dominating over $P_{\text{LA}}$ and  $P_{\text{ZA}}$  and for the large part of the higher temperature region $P_{\text{TA}}$ is greater by $\sim$$1$~order of magnitude. This may be attributed to the relatively  larger  $D_{1\text{TA}}$ ($>D_{1\text{LA}}$ and  $D_{1\text{ZA}}$) and smaller $v_{\text{TA}}$ ($<v_{\text{LA}}$). The cross over temperature between $P_{\text{TA}}$ and $P_{\text{ZA}}$ phonons  depends upon their $v_{\lambda}$ and $D_1$ values. It is also noticed  that  for  $T_e \lesssim 3$~K ($\gtrsim 30$~K), $P_{\text{LA}}$ ($P_{\text{TA}}$) in silicene is almost same as that of graphene. We emphasize that the values of $D_1$ and $v_{\lambda}$ of the respective phonon modes will play a significant role in determining their relative contribution in  different temperature region.
 
 \begin{figure}[!t]
 	\includegraphics[width=\columnwidth]{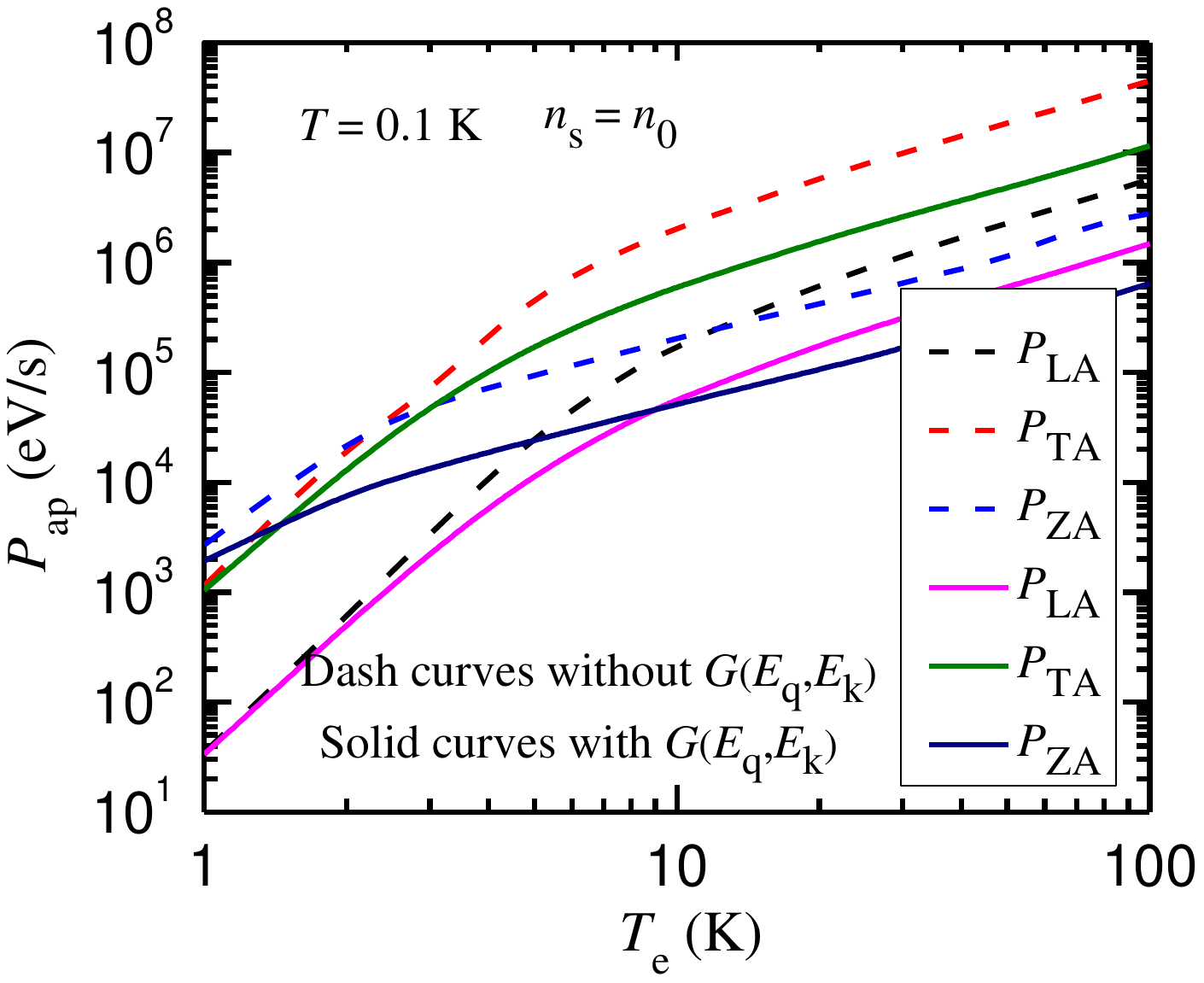}
 	\caption{Power loss $P_{ap}$ due to intravalley acoustic phonons as a function of $T_e$ with and without chiral function $G(E_q,E_k)$,  at $T = 0.1$~K.}
 	\label{fig:Fig2}
 \end{figure}
 
 The effect of chiral function $G(E_k, E_q)$ on $P_{\text{ap}}$  may be seen by  presenting it  as a function of $T_e$ with and without chiral function (Fig.~\ref{fig:Fig2}). The $P_{\text{ap}}$ with chiral function is about $2-4$ times smaller than the $P_{\text{ap}}$ without chiral function. The difference is larger at higher $T_e$. It may be recalled that the momentum relaxation time due to intravalley acoustic phonons with chiral function~\cite{prb77(2008)115449} is 4 times smaller than the one  without chiral function~\cite{prb87(2013)115418}. 
 
  \begin{figure*}[!t]
	\includegraphics[width=\textwidth]{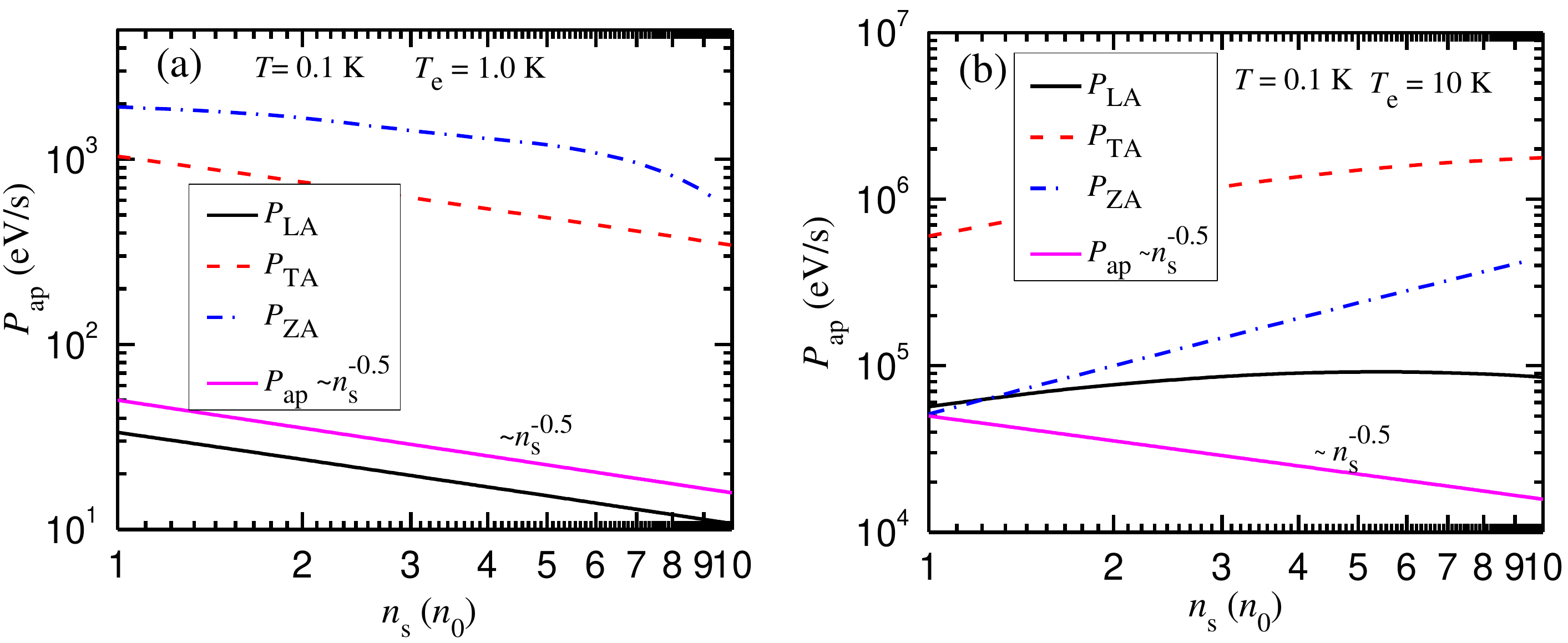}
	\caption{Power loss $P_{\text{ap}}$ due to intravalley acoustic phonons as a function of  electron density $n_s$. at $T = 0.1$~K.  (a) $T_e  = 1$~K and (b) $T_e =10$~K.}
	\label{fig:Fig3}
\end{figure*}
 
 The electron density $n_s$ dependence of intravalley $P_{\text{ap}}$ is shown in Fig.~\ref{fig:Fig3} for $T_e =1$ and $10$~K. For comparison, the curve with power law $n_s^{-1/2}$ is also shown. It is found that, at $T_e =1$~K (Fig.~\ref{fig:Fig3}(a)), behavior of $P_{\text{LA}}$ and $P_{\text{TA}}$ are as per the power law, where as $P_{\text{ZA}}$ is not as its $T_{\text{BG}}$ is very small. $P_{\text{ZA}}>P_{\text{LA}}$ and$P_{\text{TA}}$ and no cross over is found. In graphene, power law $n_s^{-1/2}$ has been experimentally observed, and has been used to identify the Dirac phase of the electron~\cite{prb87(2013)045414,jpcm27(2015)164202}. At $T_e = 10$~K (Fig.~\ref{fig:Fig3}(b)), $P_{\text{LA}}$, $P_{\text{TA}}$ and $P_{\text{ZA}}$ increase with $n_s$.  $P_{\text{TA}}$ is much greater than $P_{\text{LA}}$ and $P_{\text{ZA}}$ with cross over of $P_{\text{LA}}$ and $P_{\text{ZA}}$ at $1.2$~K. Faster increase of $P_{\text{ZA}}$ than the other two indicates its stronger dependence on $n_s$.
 
\begin{figure}[!t]
	\includegraphics[width=\columnwidth]{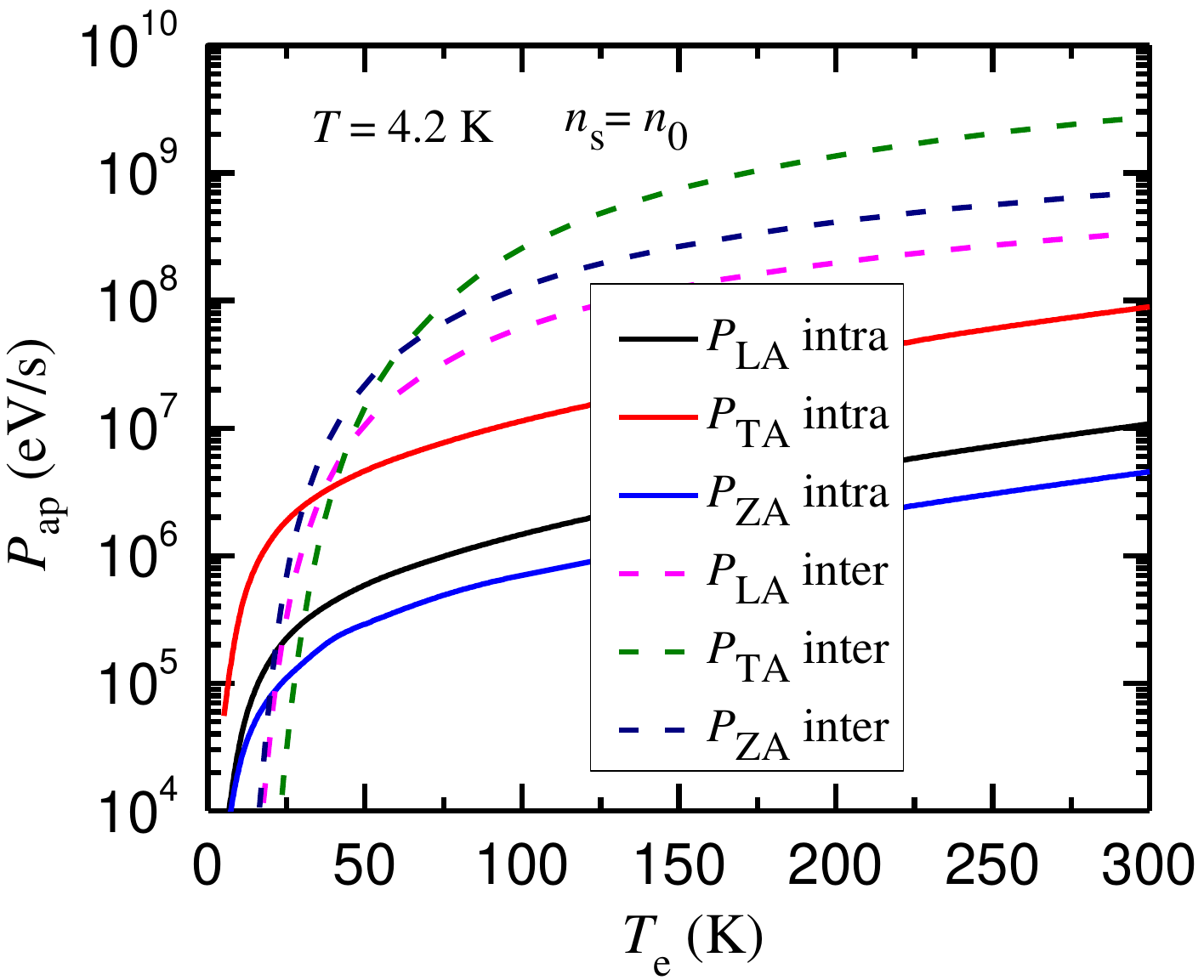}
	\caption{Power loss $P_{\text{ap}}$ due to  intravalley and intervalley  (for $\tau_p=1$~ps) acoustic phonons as a function of $T_e$  at $T = 4.2$~K.}
	\label{fig:Fig4}
\end{figure}
 
 In Fig.~\ref{fig:Fig4}, we have plotted $P_{\text{ap}}$ due to intravalley and intervalley acoustic phonons  in the range $T_e = 4.2-300$~K, for $T= 4.2$~K, taking the hot phonon effect for intervalley acoustic phonons with $\tau_p = 1$~ps. Considering the contribution to $P_{\text{ap}}$ only by the intravalley acoustic phonons, it is found that $P_{\text{ap}}$ due to TA phonons is dominant in the entire range of $T_e$, unlike the behavior given for  $T= 0.1$~K. At $T_e = 300$~K, $P_{\text{ap}}$ due to TA phonons  is about $10 (20)$ times greater than that due to LA (ZA) phonons. In the contribution to $P_{\text{ap}}$ due to intervalley acoustic phonons, for \mbox{$T_e\lesssim 50$~K},  $P$ due to ZA phonons  is predominant (because of their relatively smaller \mbox{$\hbar\omega_0=13.2$~meV} and larger $D_0 =  6.1\times 10^7$~eV/cm), whereas for \mbox{$T_e\gtrsim 50$~K}, TA phonon contribution is dominant. In both of these cases, LA phonon contribution to $P_{\text{ap}}$ is significant and it is in between (smaller than)  the  TA and ZA contributions for $T_e \lesssim 50$~K ($\gtrsim 50$~K). Comparing the $P_{\text{ap}}$ due to intravalley and intervalley acoustic phonons, the latter is beginning to dominate at relatively smaller $T_e> 20-30$~K, because of their smaller energies (see Table~\ref{table1}).  It is important to note that, the relative contribution of these intervalley modes to the power loss and their cross over depends upon the phonon energy $\hbar\omega_{0\lambda}$ and deformation potential constant $D_{0\lambda}$ of the respective modes.

\begin{figure}[!b]
	\includegraphics[width=\columnwidth]{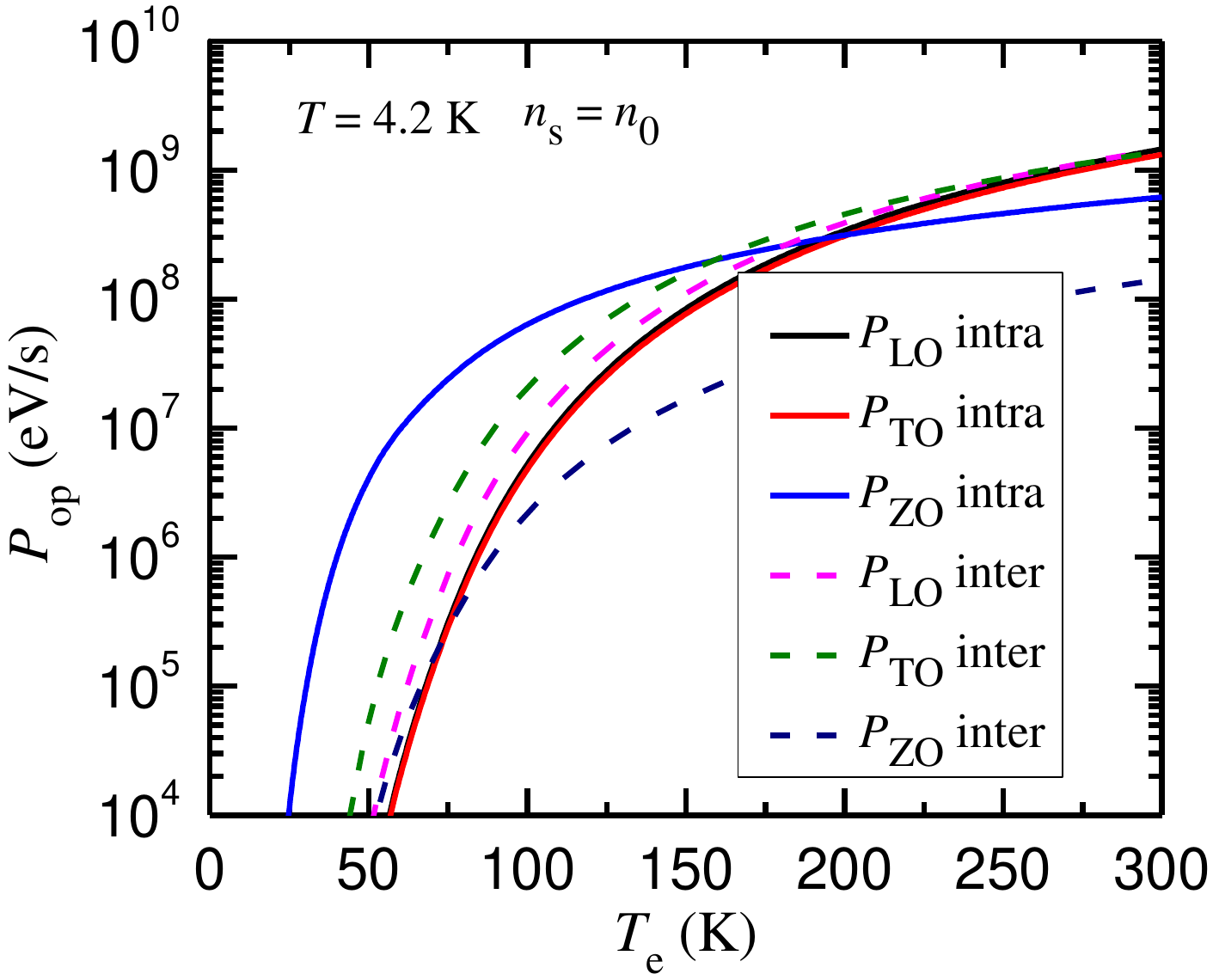}
	\caption{Power loss $P_{\text{op}}$ due to  intravalley and intervalley optical  phonons as a function of  $T_e$ at $T = 4.2$~K and for $\tau_p=1$~ps.}
	\label{fig:Fig5}
\end{figure}

The contribution to power loss $P_{\text{op}}$ by optical modes LO, TO, and ZO (both intravalley and intervalley) is shown in Fig.~\ref{fig:Fig5} for $\tau_p=1$~ps.  For $T_e \lesssim 150$~K, $P_{\text{op}}$ is governed by the intravalley ZO phonon scattering and  for $T_e \gtrsim 150$~K, $P_{\text{op}}$ due to LO and TO phonons (both intravalley and intervalley) are equally dominating ($10^9$~eV/s). The cross over  in the contributions is again determined by the respective phonon energies and their deformation potential coupling constants. The values of $P_{\text{op}}$, at $T_e=300$~K,  are  of the same order of magnitude as in graphene~\cite{prb79(2009)235406},  but relatively smaller. The $P_{\text{op}}$ in silicene is about $2$ orders of magnitude smaller than those in  monolayer MoS$_2$~\cite{prb90(2014)165436} and 3D Dirac semimetal (3DDS)~\cite{jpcm30(2018)265303}. 

\begin{figure}[!t]
	\includegraphics[width=\columnwidth]{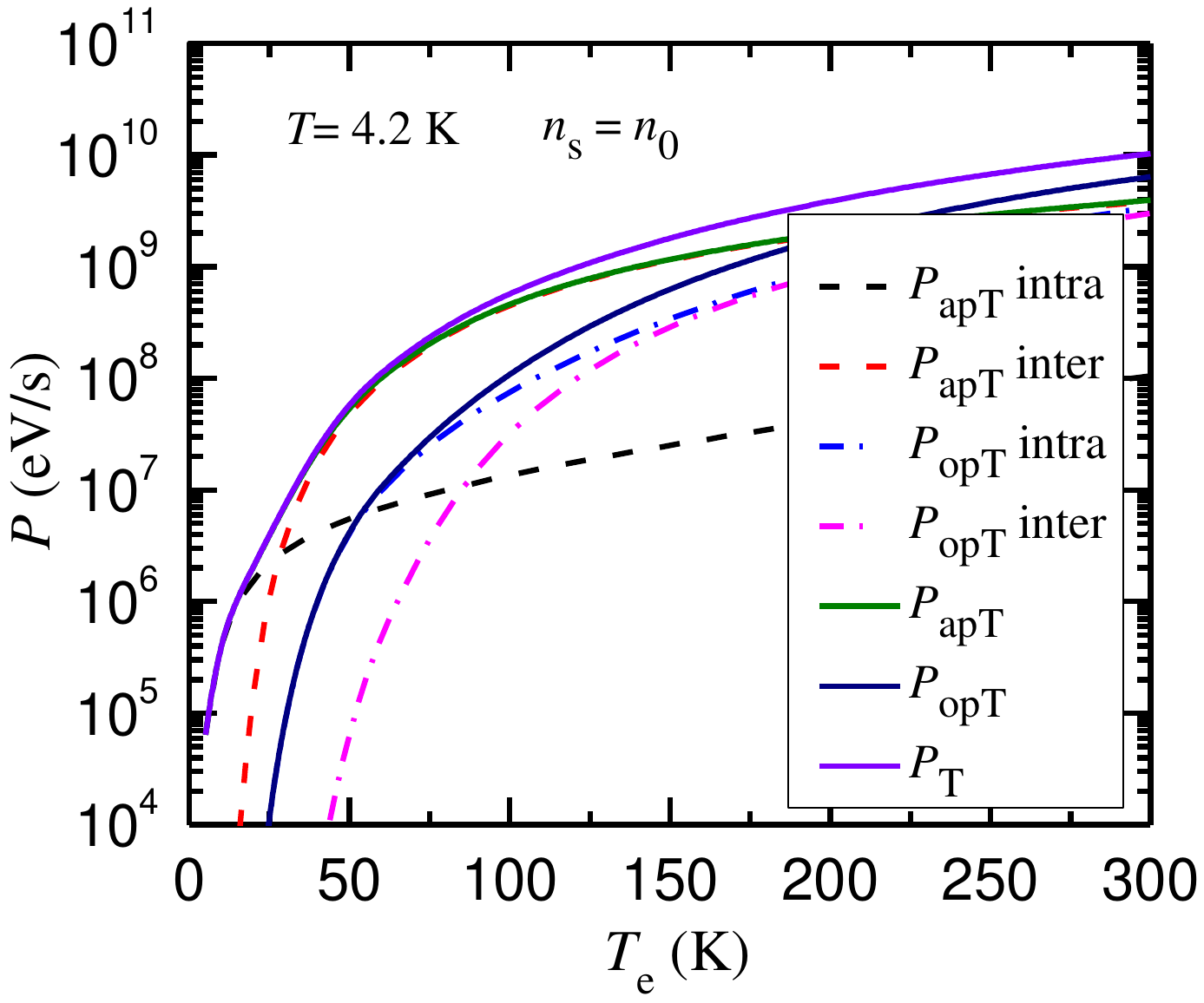}
	\caption{Power loss due to total of intravalley acoustic $P_{\text{apT}}$(intra), intervalley acoustic $P_{\text{apT}}$(inter), intravalley optical  $P_\text{opT}$(intra) and  intervalley  optical $P_\text{opT}$ (inter) phonons as a function of $T_e$ at $T = 4.2$~K and for $\tau_p=1$~ps. Curves of  totals $P_\text{apT}= P_\text{apT}\text{(intra)}+P_\text{apT}\text{(inter)}$, $P_\text{opT} = P_\text{opT}\text{(intra)}+P_\text{opT}\text{(inter)}$, and $P_{\text{T}} = P_\text{apT} + P_\text{opT}$ are also shown.}
	\label{fig:Fig6}
\end{figure}

\begin{figure}[!b]
	\includegraphics[width=\columnwidth]{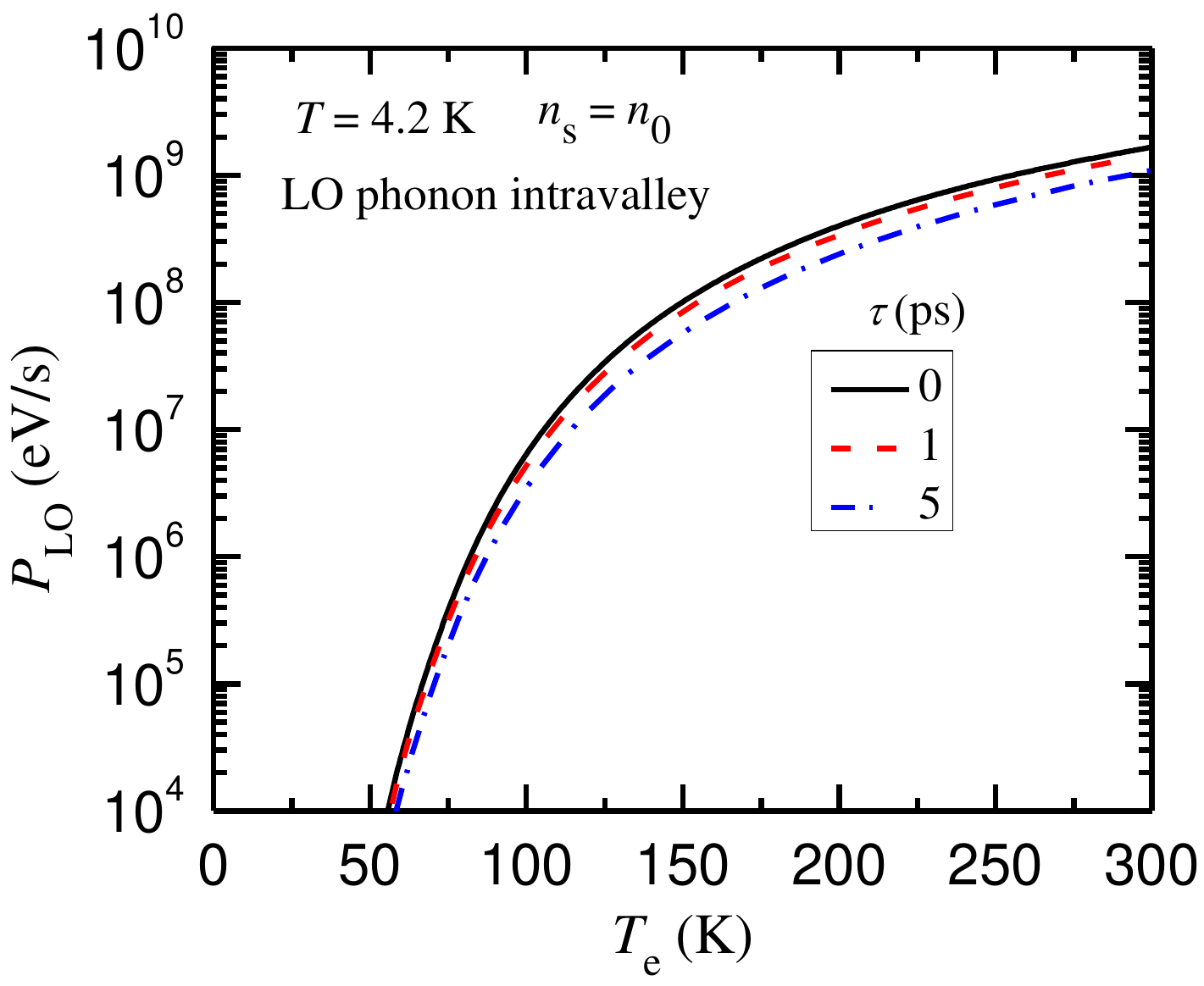}
	\caption{Power loss $P_\text{op}$ due to  intravalley LO  phonons as a function of $T_e$  for $\tau_p=0, 1$ and $5$~ps, and  $T = 4.2$~K.}
	\label{fig:Fig7}
\end{figure}

To analyze the relative contributions from the  acoustic and optical phonons, we have plotted in Fig.~\ref{fig:Fig6} the  total power loss due to acoustic intravalley $P_\text{apT}\text{(intra)}$,  acoustic intervalley $P_\text{apT}\text{(inter)}$, optical intravalley $P_\text{opT}\text{(intra)}$ and optical intervalley $P_\text{opT}\text{(inter)}$ as a function of $T_e$. These are calculated at  $T=4.2$~K, for $n_s= n_0$ and  $\tau_p=1$~ps.  Moreover, the total contribution of acoustic phonons $P_\text{apT} = P_\text{apT}\text{(intra)} + P_\text{apT}\text{(inter)}$, optical phonons $P_\text{opT}=P_\text{opT}\text{(intra)}+P_\text{opT}\text{(inter)}$  and total of acoustic and optical  $P_\text{T} = P_\text{apT} + P_\text{opT}$  are also presented. It is found that $P_\text{apT} \text{(intra)}$ is dominant for $T_e\lesssim 30$~K and $P_\text{apT}\text{(inter)}$ is dominant over all the  other mechanisms in the range $\sim30~\text{K}<T_e\lesssim 200$~K, with $P_\text{apT}\approx P_\text{apT} \text{(inter)}$. The cross over from $P_\text{apT}$ to $P_\text{opT}$ takes place at about $200$~K and above this $T_e$  the  $P_\text{opT}$ is predominant. It may be recalled that, in graphene,  significant/dominant contribution to $P$ from optical phonons comes for $T_e\gtrsim 250$~K~\cite{prb79(2009)235406}. 

In order to understand the effect of  heating of the optical phonons on power dissipation, we show in Fig.~\ref{fig:Fig7} $P_\text{op}$ due to intravalley LO~phonons as a function of  $T_e$  for $\tau_p= 0,1$ and $5$~ps. It is observed that the hot phonon effect is reducing the power loss, as expected because the phonon heating leads to reabsorption processes. However,  for the chosen $n_s$ we find that the reduction in power loss is small as found in monolayer MoS$_2$~\cite{prb90(2014)165436}, and it is still smaller for phonons of small energy (ZO). The same effect is found with the other intervalley acoustic and optical phonons. It is expected that for still larger $n_s$ ($>10 n_0$), the reduction in $P_\text{op}$ may be larger, because large $n_s$ increases hot phonon number and thereby enhancing their reabsorption. Rengel and co-workers~\cite{jpdap51(2018)415102} have shown that the hot phonon effect has less impact on drift velocity in silicene than in graphene.

\begin{figure}[!t]
	\includegraphics[width=\columnwidth]{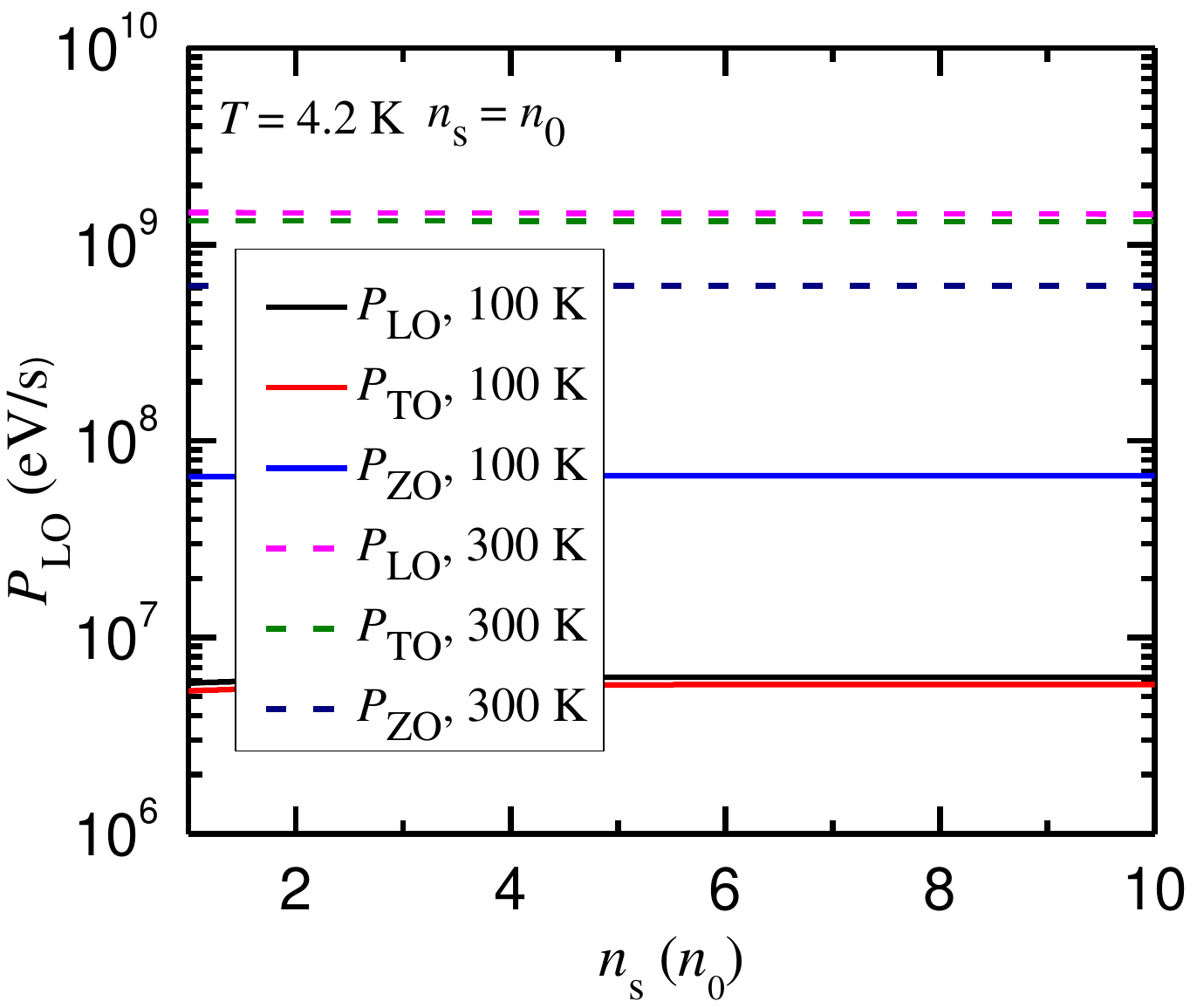}
	\caption{Power loss $P_\text{op}$ due to  intravalley optical  phonons as a function of electron density $n_s$ at $T_e  = 100$ and $300$~K, for $\tau_p=1$~ps  and $T = 4.2$~K.}
	\label{fig:Fig8}
\end{figure}

In Fig.~\ref{fig:Fig8}, $P_\text{op}$  is depicted for intravalley optical (LO, TO and ZO) phonons as a function of $n_s$ for $T_e =100$ and $300$~K and $\tau_p=1$~ps. For all the three modes, $P_\text{op}$ is found to be nearly constant in the range of $n_s =1-10$~$n_0$ considered. A similar observation is made in monolayer MoS$_2$~\cite{prb90(2014)165436} for $n_s$ up to $n_0$. We expect nearly the same $n_s$ independent behavior  of the power loss due to  intervalley acoustic and optical phonons. We point out that in graphene $P_\text{op}$ is weakly increasing with $n_s$~\cite{prb79(2009)235406}. 

\begin{figure*}[!t]
	\includegraphics[width=\textwidth]{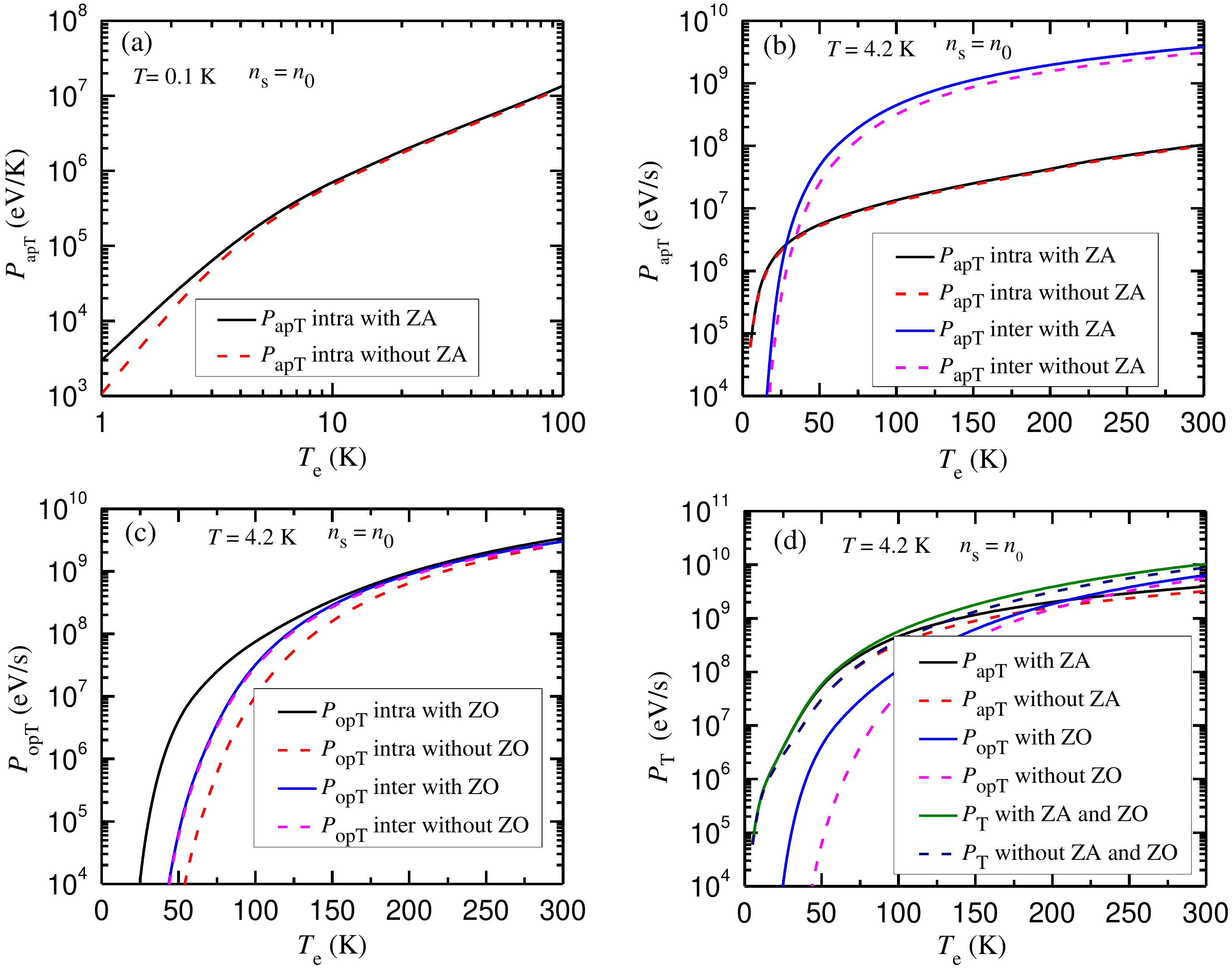}
	\caption{Total power loss due to  intravalley and intervalley  acoustic and optical phonon scattering as a function of $T_e$ with and without out-of-plane phonons ZA and ZO. (a) $P_\text{apT}$ due to  intravalley acoustic  phonons with and without ZA phonons at  $T = 0.1$~K, (b) $P_\text{apT}$ due to  intravalley and intervalley acoustic  phonons with and without ZA phonons, (c) $P_\text{opT}$ due to  intravalley and intervalley optical  phonons with and without ZO phonons and (d) Curves of  total $P_\text{apT} = P_\text{apT}\text{(intra)}+P_\text{apT}\text{(inter)}$, $P_\text{opT}= P_\text{opT}\text{(intra)}+P_\text{opT}\text{(inter)}$, and $P_\text{T} = P_\text{apT} + P_\text{opT}$. In figures (b), ( c) and (d) curves are for $\tau_p=1$~ps  and $T = 4.2$~K.}
	\label{fig:Fig9}
\end{figure*}

We examine the influence of out-of-plane modes to the power loss by plotting in Fig.~\ref{fig:Fig9} the results of the total power loss due to acoustic and optical phonons with and without considering the contribution from the out-of-plane modes (ZA and ZO). The total of intravalley acoustic phonons at $T = 0.1$~K with and without ZA phonons is presented in Fig.~\ref{fig:Fig9}(a). The total of intravalley and intervalley acoustic phonons and optical phonons are depicted in Fig.~\ref{fig:Fig9}(b) and (c), respectively, with and without the ZA and ZO phonon contributions.  Finally, the total of all the acoustic and optical modes with and without the contribution of ZA and ZO phonons are depicted in Fig.~\ref{fig:Fig9}(d). In Figs.~\ref{fig:Fig9}(b), (c) and (d) the power loss calculated is at $T= 4.2$~K. It is found from Fig.~\ref{fig:Fig9}(a) that, intravalley ZA phonon contribution is significant for $T_e\lesssim 4$~K, and becomes increasingly important as $T_e$ decreases further. However, the intervalley ZA phonon contribution is marginally significant in the entire range of $T_e$ (Fig.~\ref{fig:Fig9}(b)). Of all the optical modes, intravalley ZO phonon is largely responsible for the power loss (Fig.~\ref{fig:Fig9}(c)), in particular for  $T_e\lesssim 150$~K.  It is seen from Fig.~\ref{fig:Fig9}(d), for $T_e\lesssim 200$~K,  the contribution to power loss due to acoustic phonons (intravalley and intervalley) is dominating, whereas for $T_e\gtrsim 200$~K power loss by optical phonons is found to be marginally greater. Overall, it is seen that (Fig.~\ref{fig:Fig9}(d)), the addition of power dissipation due to out-of-plane (ZA and ZO) modes to the total power loss makes a small difference. 

In the following, we compare our calculations of total power loss (i.e. $P_\text{T}$ from Fig.~\ref{fig:Fig6}) with the experimental  observations of Baker {\it et al.}  in graphene with $n_s\simeq 1-1.7 n_{s0}$~\cite{prb85(2012)115403,prb87(2013)045414}  in the range $T_e = 5-100$~K, noting weak dependence on $n_s$. Interestingly, the total power dissipated to the intrinsic phonons in silicene is closer to the value experimentally observed in graphene.  For instance, for $T_e \simeq 20~(60)$~K, Baker {\it et al.}~\cite{prb85(2012)115403} have  observed $P\simeq 6.0\times 10^{-13}~(3.6\times10^{-11})$~W, while our calculations in silicene give $3.3\times10^{-13}~(1.9\times10^{-11})$~W.  However, at lower temperature  our calculations give higher  values  than the experimental results, because of the  large contribution from TA and ZA phonons than the LA phonons. For example, at about $5$~K, experimental $P\simeq 2\times 10^{-15}$~W~\cite{prb87(2013)045414}, where as our calculations give $1.0\times10^{-14}$~W.

\begin{figure}[!t]
	\includegraphics[width=\columnwidth]{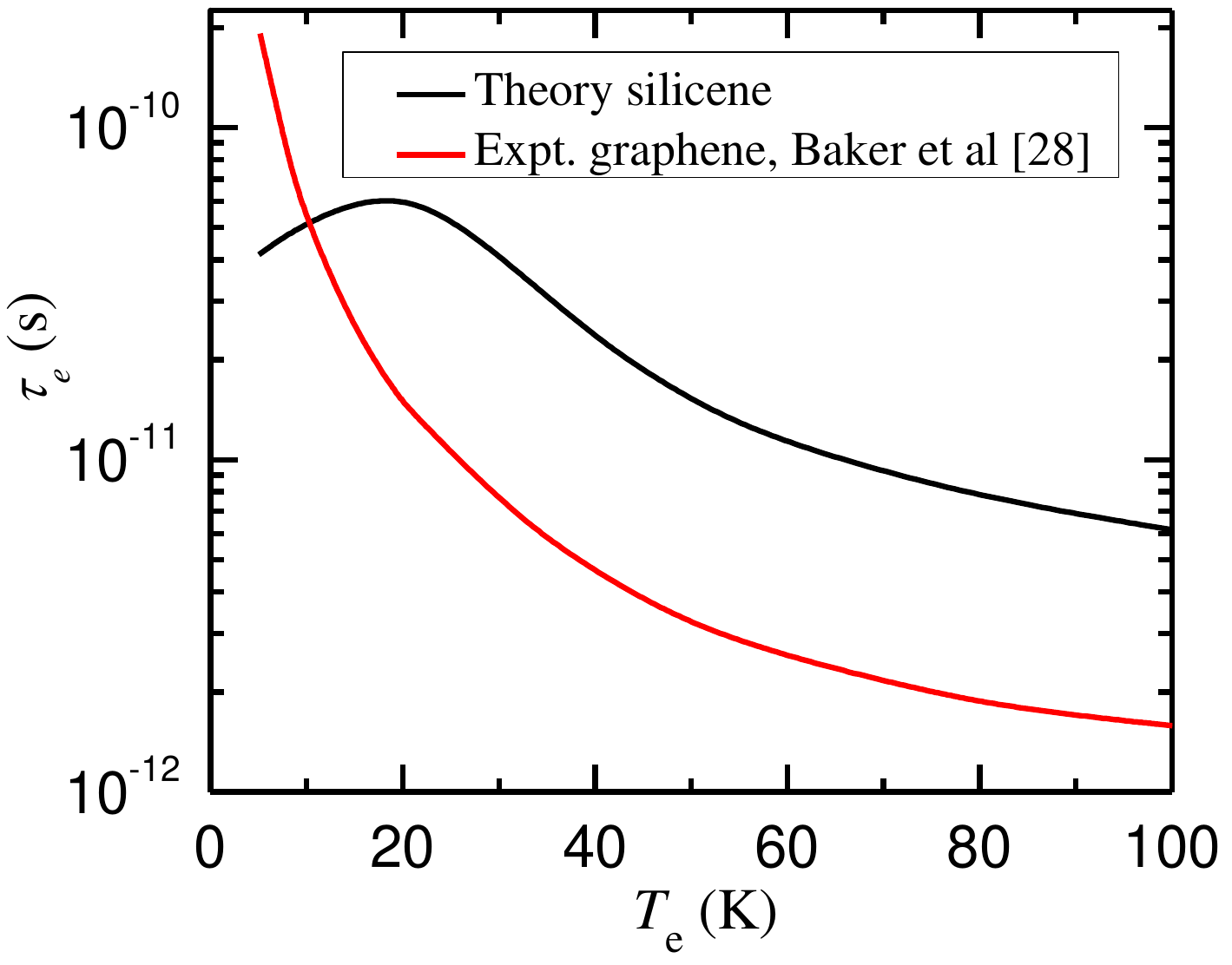}
	\caption{Energy relaxation time $\tau_e$, obtained by using total power loss $P_\text{T}$  of Fig.~\ref{fig:Fig6}, as function of $T_e$  is compared with the curve fitting to the experimental data of graphene~\cite{prb85(2012)115403}.}
	\label{fig:Fig10}
\end{figure}

The energy relaxation time $\tau_e= (p +1)(\pi k_B)^2(T_e^2-T^2)/(6E_FP)$, $p$ being the exponent of the energy dependence of the density of states, is calculated in silicene using the total $P_\text{T}$ of our calculations (Fig.~\ref{fig:Fig6}). In Fig.~\ref{fig:Fig10}, the $T_e$ dependence of $\tau_e$ is plotted together with the curve fitting to the experimental data of Baker et al in graphene~\cite{prb85(2012)115403}. For  $T_e =20-100$~K, $\tau_e$ is found to be about 4 times greater than that in graphene, and decreases monotonously with increasing $T_e$. This may be attributed to $v_F$ and $P$ both about $2$ times smaller in silicene than in graphene. For this reason, silicene may have edge over graphene for its applications in bolometers and calorimeters because of its larger value of $\tau_e$. Interestingly, $\tau_e$ exhibits a peak at $T_e $$\sim$$20$~K and decreases with decreasing $T_e$. It may be recalled that, this is the region ($T_e < 20$~K) in which $P_\text{ap}$(intra), particularly $P_\text{TA}\text{(intra)}\gg P_\text{LA}$  of graphene and silicene (see Fig.~\ref{fig:Fig1}), limits the total $P_\text{T}$.

\subsection{Power loss due to surface optical  phonons}
\begin{figure}[!b]
	\includegraphics[width=\columnwidth]{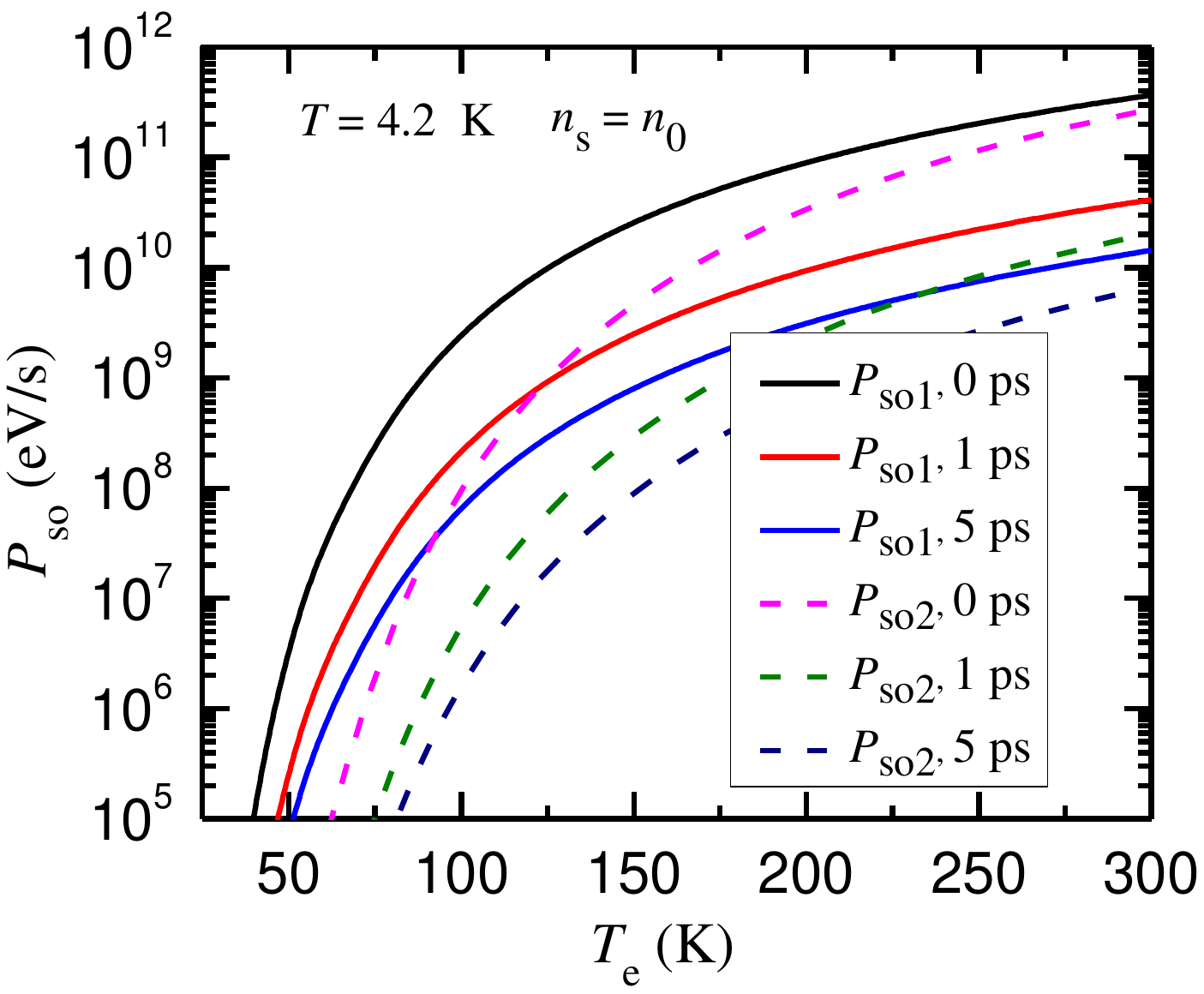}
	\caption{Power loss $P_{\text{SO}}$ due to surface optical phonons SO1 and SO2 as a function of $T_e$, with screening, for  $\tau_p= 1$ and $5$~ps, $d = 0.5$~nm, and $T = 4.2$~K.}
	\label{fig:Fig11}
\end{figure}

The power dissipation $P_{{\rm SO1}}$ and $P_{{\rm SO2}}$ due to  the surface optical modes SO1 and SO2, respectively,  of the substrate Al$_2$O$_3$ are shown in Fig.~\ref{fig:Fig11} for phonon relaxation times $\tau_p = 0,1$ and $5$~ps and $T = 4.2$~K.  The contribution from $P_{{\rm SO1}}$ is dominating in the entire range of $T_e$. This is due to the fact that {\color{blue}the} energy of SO2 ($94.29$~meV) mode is much larger than that of the SO1 ($55.01$~meV)  mode, and the probability of the emission of the former is much smaller. Hot phonon effect is found to significantly reduce the power loss, unlike the intrinsic phonons. For example, at $T_e = 300$~K, both  $P_{\rm SO1}$ and $P_{\rm SO2}$ reduce by about an order of magnitude when $\tau_p$ is changed from  $0$ to $1$~ps. {\color{blue}A} similar large hot phonon effect is found in 3DDS, in which electron interaction with optical phonons is via Fr\"{o}hlich coupling~\cite{jpcm30(2018)265303}. 

\begin{figure}[!t]
	\includegraphics[width=\columnwidth]{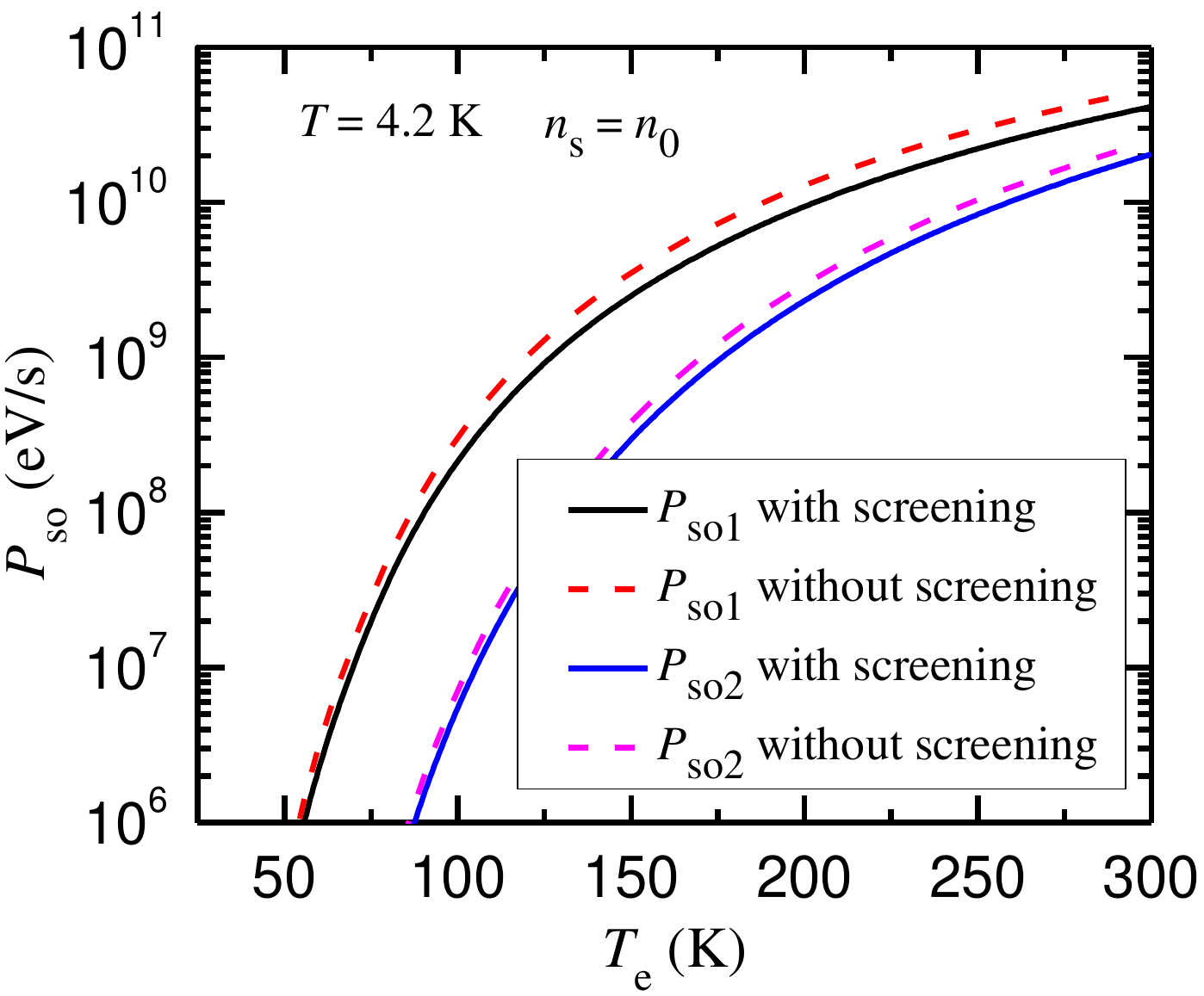}
	\caption{Power loss $P_{\text{SO}}$ due to SO1 and SO2 phonons as a function of $T_e$, with and without screening for  $\tau_p= 0, 1$~ps, $d = 0.5$~nm,  and $T = 4.2$~K.}
	\label{fig:Fig12}
\end{figure}

The effect of screening is expected to degrade the power loss $P_{\text{SO}}$, however, it is found from Fig.~\ref{fig:Fig12} that the reduction is small. The effect of screening occurs in two places in the power loss. Because of the reduced interaction by screening, the hot phonon number is reduced, which will enhance the power loss. On the other hand, the power loss which is directly depending on electron-SO phonon coupling strength may be reduced by screening. These two causes may lead to the reduced effect of screening on the power loss. 

\begin{figure}[!b]
	\includegraphics[width=\columnwidth]{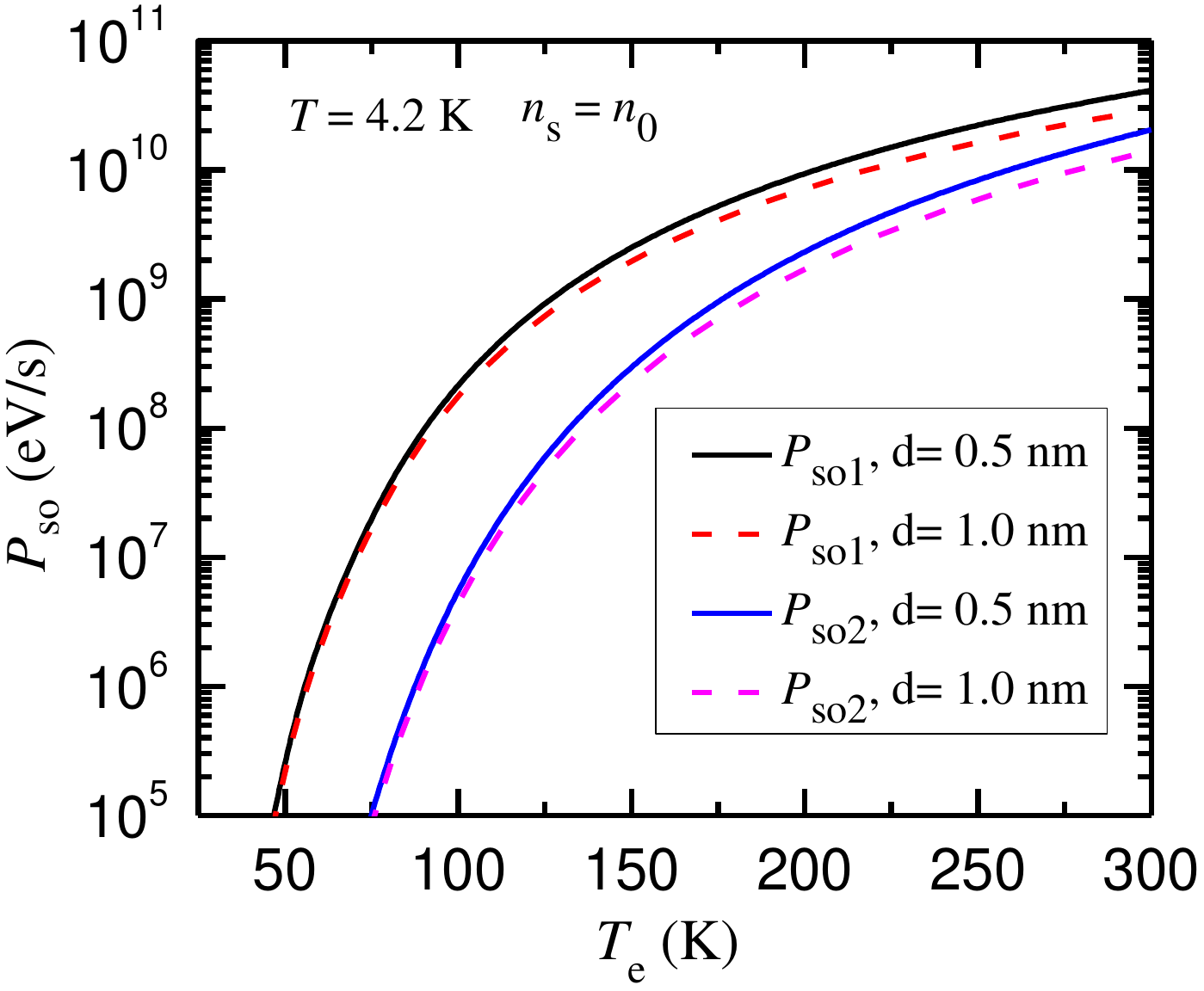}
	\caption{Power loss $P_{\text{SO}}$ due to SO1 and SO2 phonons as a function of $T_e$, with  screening for  $d = 0.5$ and $1.0$~nm,  $\tau_p=1$~ps and  $T = 4.2$~K.}
	\label{fig:Fig13}
\end{figure}

Transport studies in graphene on the substrate are carried out for different values of distance $d~$($\sim$$0.2- 0.8$~nm) between the substrate and graphene~\cite{prb82(2010)115452}. In order to see the effect of $d$ in silicene, $P_{\text{SO}}$ is presented in Fig.~\ref{fig:Fig13} for $d = 0.5$ and $1.0$~nm. With the increasing separation, power dissipation is expected to degrade due to the reduced interaction. However, as discussed in the case of screening, increased separation reduces not only hot phonon numbers but also electron-SO phonon coupling strength, resulting in a smaller effect of $d$ on the power loss. 

\begin{figure}[!t]
	\includegraphics[width=\columnwidth]{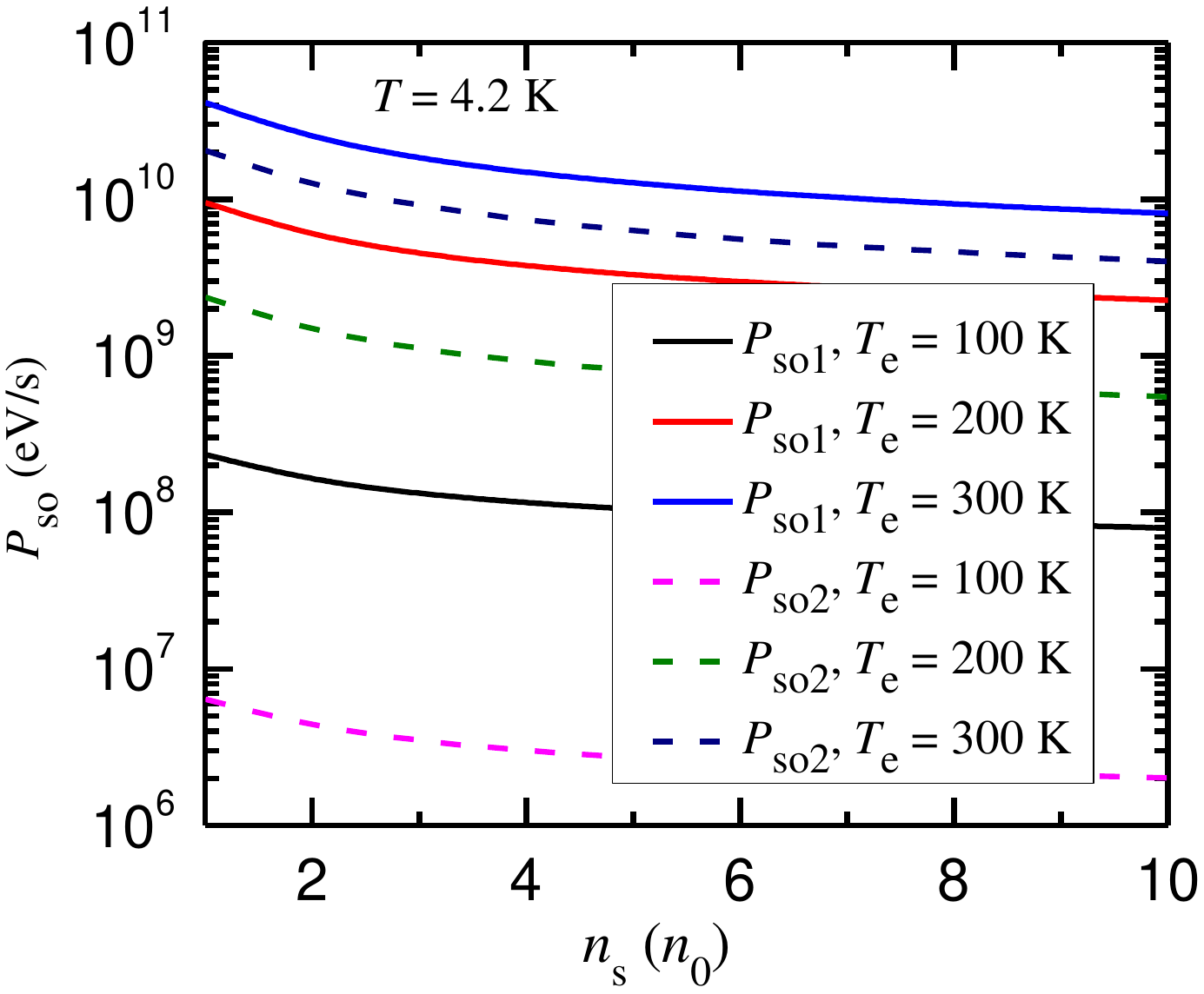}
	\caption{Power loss $P_\text{SO}$ due to phonons SO1 and SO2 phonons as a function of $n_s$, with  screening,  for $T_e=100$, $200$, and $300$~K, $d = 0.5$~nm, $\tau_p= 1$~ps at $T = 4.2$~K.}
	\label{fig:Fig14}
\end{figure}

In Fig.~\ref{fig:Fig14}, $P_{\text{SO}}$ is shown as a function of $n_s$ for \mbox{$T_e = 100$}, $200$ and $300$~K. At all these temperatures, $P_{\text{SO1}}$ and $P_{\text{SO2}}$ decrease with increase of $n_s$. When $n_s$ increases by an order of magnitude, both $P_{\text{SO1}}$ and $P_{\text{SO2}}$ reduce by a factor $\sim$$4$, which is comparable to the observation made  in bilayer graphene~\cite{jap113(2013)063705}.

\begin{figure}[!h]
	\centering
	\includegraphics[width=\columnwidth]{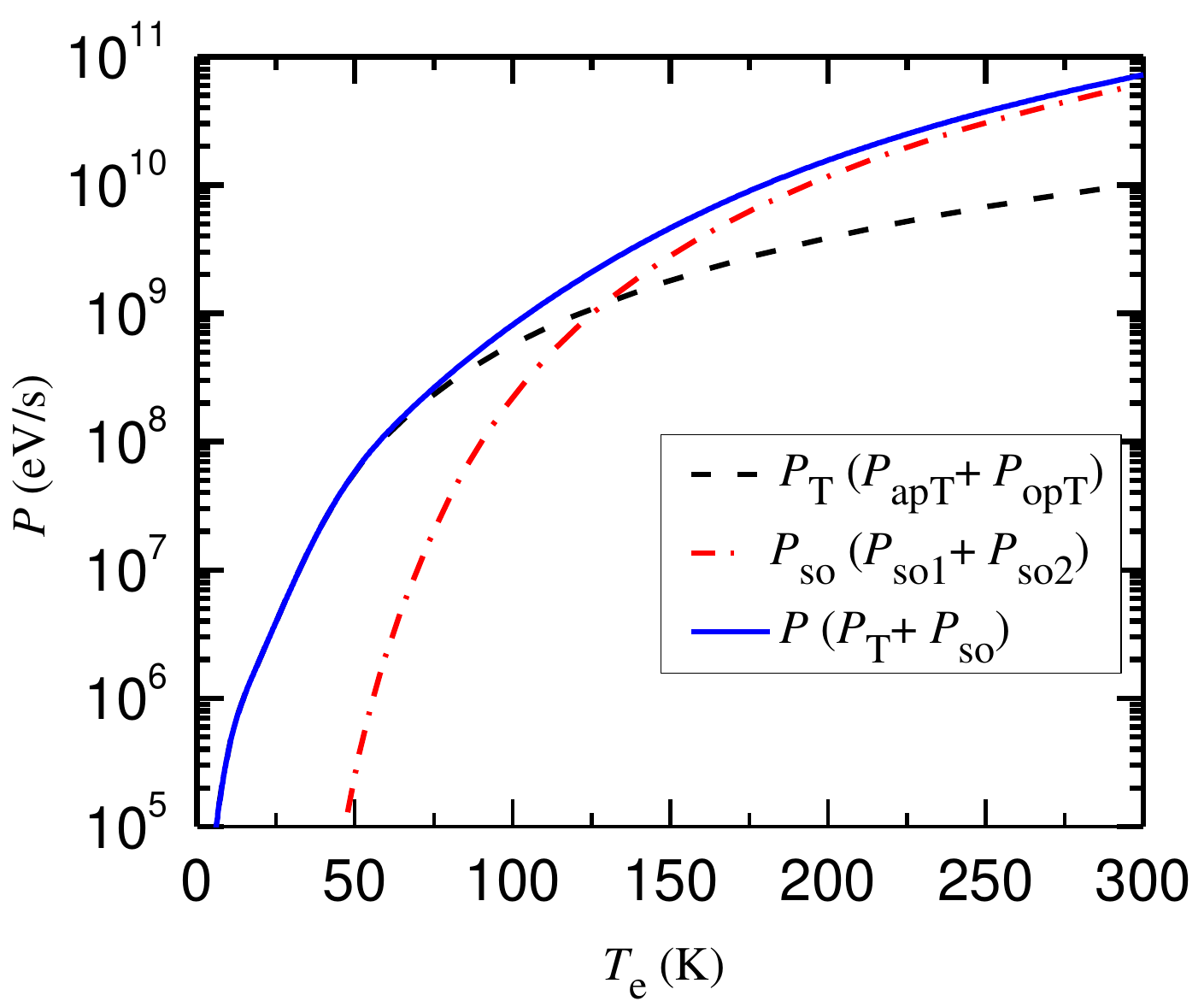}
	\caption{Power loss $P$ total of all intrinsic acoustic and optical phonons and surface optical phonons as a function of $T_e$ for $n_s = n_0$, $\tau_p= 1$~ps , $d  = 0.5$~nm at $T = 4.2$~K.}
	\label{fig:Fig15}
\end{figure}

Finally, we have presented in Fig.~\ref{fig:Fig15} the total power loss due to all intrinsic acoustic and  optical phonons $P_T$ (from Fig.~\ref{fig:Fig6}) and the total of surface optical phonons $P_{\rm SO} (=P_{\rm SO1}  + P_{\rm SO2})$ as a function of $T_e$,  for $n_s = n_0$, $\tau_p= 1$~ps at $T= 4.2$~K. $P_{\rm SO}$ is taken for $d = 0.5$~nm with screening. The cross over of the $P_{\rm SO}$ and $P_{\rm T}$ takes place at $T_e \simeq 130$~K, above which$P_{\rm SO}$ is predominant. Further, the difference keeps on increasing with $P_{\rm SO}$ being nearly $6$ times the $P_{\rm T}$ at $T_e = 300$~K.

We would like to make the following remarks with regard to silicene on the substrate.  Yeoh {\it et al.}~\cite{sst31(2016)065012} argue that one of the reasons for the degradation of mobility in the supported sample may be renormalization of $v_F$, attributing to the interaction of Al$_2$O$_3$ with silicene resulting in a structural reconstruction which renormalizes the silicene band structure~\cite{prb94(2016)075409}. In Ref.~\cite{sst31(2016)065012} the deformation potential coupling is applied for SO phonon interaction with the electrons in silicene and make the choice of $v_F = 3.4\times 10^7$~cm/s to match the low field room temperature mobility of $100$~cm$^2$/Vs~\cite{nnano10(2015)227}. Whereas, for polar substrates, electron coupling with surface modes is via  Fr\"{o}hlich coupling~\cite{prb86(2012)045413,prb82(2010)115452}, which is adopted in the present study. In Ref.~\cite{jms1199(2020)126878} with Fr\"{o}hlich coupling and $v_F= 3.4\times10^7$~cm/s, authors have obtained a mobility close to the experimental value.

In the present analysis, SO phonon contribution to the power loss is significantly large, because of the strong Fr\"{o}hlich coupling,  compared to the intrinsic phonons. The effect of  SO phonon scattering can be reduced by ‘substrate engineering’. The choice of a polar substrate with the appropriate SO phonon energies and high- and low- frequency dielectric constants, may be one possible way to reduce the substrate effect. For instance, Konar {\it et al.}~\cite{prb82(2010)115452} suggest the dielectrics, with intermediate dielectric constants, such as AlN and SiC, as the optimum choice for gate insulators for graphene.  It is also suggested in graphene that heat dissipation through SO phonons can be suppressed through a non-polar substrate such as diamond-like carbon~\cite{prb86(2012)045413} so that $\tau_e$ can be enhanced. Samples with slower power transfer to the substrate are preferred for applications in bolometers and calorimeters. 

It is important to note that, in the literature, the expressions given for momentum relaxation time $\tau_m$  due to acoustic deformation potential coupling are differing by a constant factor 2 or 4~\cite{prb87(2013)115418,sst31(2016)065012,sst31(2016)115004,sst33(2018)065011,prb81(2010)195442,jpcm21(2009)232204}. We could obtain  Eq.~(8) of Hwang and Das Sarma~\cite{prb77(2008)115449} by taking the angular dependence of the electron-acoustic phonon matrix element $(1+\cos\theta)/2$, due to chiral nature of the wave function, and $(1-\cos\theta)$ coming from the definition of relaxation time for elastic scattering. It is suspected that these differences may be due to the difference in the angular dependence of the el-ph matrix element, the difference in the definition of the scattering matrix element, and the confusion between scattering probability and the inverse relaxation time~\cite{prb81(2010)195442,Rengel}. These differences in factor/s in turn lead to a discrepancy in $D_1$ values used by various authors.  Moreover, we also note that to get an agreement with the results of the FMC model, the authors in Refs.~\cite{sst31(2016)065012,sst31(2016)115004,sst33(2018)065011} have varied deformation potential coupling constants in their AFC model. It is also found that the results of Li {\it et al.}~\cite{prb87(2013)115418} and Fischetti and co-workers~\cite{jap124(2018)044306} also differ. Fischetti~\cite{Fischetti} suspects that the difference is due to the different discretization mesh of $k$-points used to calculate the electron-phonon matrix elements. We believe experimental data will be of great help to set these parameters. 

In the power loss calculations due to LO and TO  phonon scattering in graphene, the angular dependence of the electron-optical phonon matrix element for these modes are, respectively, $(1-\cos\theta)/2$ and $(1+\cos\theta)/2$,  and became negated after summing~\cite{prb86(2012)045413}.  However, the angular dependencies for all the intervalley acoustic phonons, intravalley and intervalley optical phonon scattering in silicene are not known. 

Finally, we would like to mention that our investigation of the power dissipation of Dirac fermions is fully analytical and will be of great help to interpret the experimental observations. In the transport study, it is pointed out that the analytical model may provide an intuitive interpretation of the observed properties by means of basic phenomena, even if they give less accurate results~\cite{sst31(2016)115004,prb93(2016)035414}. The contribution of ZA phonons with quadratic dispersion to the power dissipation, due to the buckled nature of the silicene, and its suppression may require separate addressing. 

\section{Conclusions}
Hot electron power loss $P$ is  analytically studied in suspended silicene by considering the electron scattering by intra and intervalley acoustic (LA, TA, and ZA) and optical (LO, TO, and ZO) phonons using the phonon energies and deformation potential constants extracted by Li {\it et al.}~\cite{prb87(2013)115418} from FMC model.  For electron temperature $T_e\lesssim30$~K, the total power loss is governed by the intravalley acoustic phonons, particularly by TA phonons, whereas for $\sim$$30$~K$<T_e\lesssim 200$~K, intervalley acoustic phonons are the dominant cooling channels. Above $200$~K, heat transfer is predominantly by the intra and intervalley optical phonons. The total power loss in silicene is of the same order of magnitude (but smaller)  as in graphene. However, for $T_e\lesssim 20~$($\sim$$2$~K), where TA (ZA) phonons are the dominant mechanisms, the total $P$ in silicene is greater than that in graphene by an order of magnitude. At not too low $T_e$, $P$ due to intravalley acoustic phonons increases with electron density $n_s$. On the other hand, $P$ due to the intervalley acoustic and intra and intervalley optical phonons is found to be independent of $n_s$. The energy relaxation time in silicene is found to be nearly 4 times greater than that in graphene and hence it may find applications in bolometers and THz detectors.

Significance of the scattering by out-of-plane (ZA and ZO) modes is examined by calculating  $P$ with and without these phonons. Their contribution to the total $P$ is marginally significant.

In silicene on the substrate Al$_2$O$_3$, the power transfer, $P_{\text{SO}}$, to the SO phonons of the substrate is found to be greater ($\sim$$6$ times) than the intrinsic phonons in the higher $T_e$ region. $P_{\text{SO}}$ is decreasing with increasing $n_s$ and the hot phonon effect is found to reduce this power loss significantly. The possibility of suppression of $P_{\text{SO}}$ is discussed.



\begin{thebibliography}{42}%
	\makeatletter
	\providecommand \@ifxundefined [1]{%
		\@ifx{#1\undefined}
	}%
	\providecommand \@ifnum [1]{%
		\ifnum #1\expandafter \@firstoftwo
		\else \expandafter \@secondoftwo
		\fi
	}%
	\providecommand \@ifx [1]{%
		\ifx #1\expandafter \@firstoftwo
		\else \expandafter \@secondoftwo
		\fi
	}%
	\providecommand \natexlab [1]{#1}%
	\providecommand \enquote  [1]{``#1''}%
	\providecommand \bibnamefont  [1]{#1}%
	\providecommand \bibfnamefont [1]{#1}%
	\providecommand \citenamefont [1]{#1}%
	\providecommand \href@noop [0]{\@secondoftwo}%
	\providecommand \href [0]{\begingroup \@sanitize@url \@href}%
	\providecommand \@href[1]{\@@startlink{#1}\@@href}%
	\providecommand \@@href[1]{\endgroup#1\@@endlink}%
	\providecommand \@sanitize@url [0]{\catcode `\\12\catcode `\$12\catcode
		`\&12\catcode `\#12\catcode `\^12\catcode `\_12\catcode `\%12\relax}%
	\providecommand \@@startlink[1]{}%
	\providecommand \@@endlink[0]{}%
	\providecommand \url  [0]{\begingroup\@sanitize@url \@url }%
	\providecommand \@url [1]{\endgroup\@href {#1}{\urlprefix }}%
	\providecommand \urlprefix  [0]{URL }%
	\providecommand \Eprint [0]{\href }%
	\providecommand \doibase [0]{http://dx.doi.org/}%
	\providecommand \selectlanguage [0]{\@gobble}%
	\providecommand \bibinfo  [0]{\@secondoftwo}%
	\providecommand \bibfield  [0]{\@secondoftwo}%
	\providecommand \translation [1]{[#1]}%
	\providecommand \BibitemOpen [0]{}%
	\providecommand \bibitemStop [0]{}%
	\providecommand \bibitemNoStop [0]{.\EOS\space}%
	\providecommand \EOS [0]{\spacefactor3000\relax}%
	\providecommand \BibitemShut  [1]{\csname bibitem#1\endcsname}%
	\let\auto@bib@innerbib\@empty
	\bibitem [{\citenamefont {Cahangirov}\ \emph {et~al.}(2009)\citenamefont
		{Cahangirov}, \citenamefont {Topsakal}, \citenamefont {Akt\"urk},
		\citenamefont {Sahin},\ and\ \citenamefont {Ciraci}}]{prl102(2009)236804}%
	\BibitemOpen
	\bibfield  {author} {\bibinfo {author} {\bibfnamefont {S.}~\bibnamefont
			{Cahangirov}}, \bibinfo {author} {\bibfnamefont {M.}~\bibnamefont
			{Topsakal}}, \bibinfo {author} {\bibfnamefont {E.}~\bibnamefont {Akt\"urk}},
		\bibinfo {author} {\bibfnamefont {H.}~\bibnamefont {Sahin}}, \ and\ \bibinfo
		{author} {\bibfnamefont {S.}~\bibnamefont {Ciraci}},\ }\href {\doibase
		10.1103/PhysRevLett.102.236804} {\bibfield  {journal} {\bibinfo  {journal}
			{Phys. Rev. Lett.}\ }\textbf {\bibinfo {volume} {102}},\ \bibinfo {pages}
		{236804} (\bibinfo {year} {2009})}\BibitemShut {NoStop}%
	\bibitem [{\citenamefont {Lalmi}\ \emph {et~al.}(2010)\citenamefont {Lalmi},
		\citenamefont {Oughaddou}, \citenamefont {Enriquez}, \citenamefont {Kara},
		\citenamefont {Vizzini}, \citenamefont {Ealet},\ and\ \citenamefont
		{Aufray}}]{apl97(2010)223109}%
	\BibitemOpen
	\bibfield  {author} {\bibinfo {author} {\bibfnamefont {B.}~\bibnamefont
			{Lalmi}}, \bibinfo {author} {\bibfnamefont {H.}~\bibnamefont {Oughaddou}},
		\bibinfo {author} {\bibfnamefont {H.}~\bibnamefont {Enriquez}}, \bibinfo
		{author} {\bibfnamefont {A.}~\bibnamefont {Kara}}, \bibinfo {author}
		{\bibfnamefont {S.}~\bibnamefont {Vizzini}}, \bibinfo {author} {\bibfnamefont
			{B.}~\bibnamefont {Ealet}}, \ and\ \bibinfo {author} {\bibfnamefont
			{B.}~\bibnamefont {Aufray}},\ }\href {\doibase 10.1063/1.3524215} {\bibfield
		{journal} {\bibinfo  {journal} {Appl. Phys. Lett.}\ }\textbf {\bibinfo
			{volume} {97}},\ \bibinfo {pages} {223109} (\bibinfo {year}
		{2010})}\BibitemShut {NoStop}%
	\bibitem [{\citenamefont {Liu}, \citenamefont {Feng},\ and\ \citenamefont
		{Yao}(2011)}]{prl107(2011)076802}%
	\BibitemOpen
	\bibfield  {author} {\bibinfo {author} {\bibfnamefont {C.-C.}\ \bibnamefont
			{Liu}}, \bibinfo {author} {\bibfnamefont {W.}~\bibnamefont {Feng}}, \ and\
		\bibinfo {author} {\bibfnamefont {Y.}~\bibnamefont {Yao}},\ }\href {\doibase
		10.1103/PhysRevLett.107.076802} {\bibfield  {journal} {\bibinfo  {journal}
			{Phys. Rev. Lett.}\ }\textbf {\bibinfo {volume} {107}},\ \bibinfo {pages}
		{076802} (\bibinfo {year} {2011})}\BibitemShut {NoStop}%
	\bibitem [{\citenamefont {Drummond}, \citenamefont {Z\'olyomi},\ and\
		\citenamefont {Fal'ko}(2012)}]{prb85(2012)075423}%
	\BibitemOpen
	\bibfield  {author} {\bibinfo {author} {\bibfnamefont {N.~D.}\ \bibnamefont
			{Drummond}}, \bibinfo {author} {\bibfnamefont {V.}~\bibnamefont {Z\'olyomi}},
		\ and\ \bibinfo {author} {\bibfnamefont {V.~I.}\ \bibnamefont {Fal'ko}},\
	}\href {\doibase 10.1103/PhysRevB.85.075423} {\bibfield  {journal} {\bibinfo
			{journal} {Phys. Rev. B}\ }\textbf {\bibinfo {volume} {85}},\ \bibinfo
		{pages} {075423} (\bibinfo {year} {2012})}\BibitemShut {NoStop}%
	\bibitem [{\citenamefont {Feng}\ \emph {et~al.}(2012)\citenamefont {Feng},
		\citenamefont {Ding}, \citenamefont {Meng}, \citenamefont {Yao},
		\citenamefont {He}, \citenamefont {Cheng}, \citenamefont {Chen},\ and\
		\citenamefont {Wu}}]{nl12(2012)3507}%
	\BibitemOpen
	\bibfield  {author} {\bibinfo {author} {\bibfnamefont {B.}~\bibnamefont
			{Feng}}, \bibinfo {author} {\bibfnamefont {Z.}~\bibnamefont {Ding}}, \bibinfo
		{author} {\bibfnamefont {S.}~\bibnamefont {Meng}}, \bibinfo {author}
		{\bibfnamefont {Y.}~\bibnamefont {Yao}}, \bibinfo {author} {\bibfnamefont
			{X.}~\bibnamefont {He}}, \bibinfo {author} {\bibfnamefont {P.}~\bibnamefont
			{Cheng}}, \bibinfo {author} {\bibfnamefont {L.}~\bibnamefont {Chen}}, \ and\
		\bibinfo {author} {\bibfnamefont {K.}~\bibnamefont {Wu}},\ }\href {\doibase
		10.1021/nl301047g} {\bibfield  {journal} {\bibinfo  {journal} {Nano Lett.}\
		}\textbf {\bibinfo {volume} {12}},\ \bibinfo {pages} {3507} (\bibinfo {year}
		{2012})}\BibitemShut {NoStop}%
	\bibitem [{\citenamefont {Vogt}\ \emph {et~al.}(2012)\citenamefont {Vogt},
		\citenamefont {De~Padova}, \citenamefont {Quaresima}, \citenamefont {Avila},
		\citenamefont {Frantzeskakis}, \citenamefont {Asensio}, \citenamefont
		{Resta}, \citenamefont {Ealet},\ and\ \citenamefont
		{Le~Lay}}]{prl108(2012)155501}%
	\BibitemOpen
	\bibfield  {author} {\bibinfo {author} {\bibfnamefont {P.}~\bibnamefont
			{Vogt}}, \bibinfo {author} {\bibfnamefont {P.}~\bibnamefont {De~Padova}},
		\bibinfo {author} {\bibfnamefont {C.}~\bibnamefont {Quaresima}}, \bibinfo
		{author} {\bibfnamefont {J.}~\bibnamefont {Avila}}, \bibinfo {author}
		{\bibfnamefont {E.}~\bibnamefont {Frantzeskakis}}, \bibinfo {author}
		{\bibfnamefont {M.~C.}\ \bibnamefont {Asensio}}, \bibinfo {author}
		{\bibfnamefont {A.}~\bibnamefont {Resta}}, \bibinfo {author} {\bibfnamefont
			{B.}~\bibnamefont {Ealet}}, \ and\ \bibinfo {author} {\bibfnamefont
			{G.}~\bibnamefont {Le~Lay}},\ }\href {\doibase
		10.1103/PhysRevLett.108.155501} {\bibfield  {journal} {\bibinfo  {journal}
			{Phys. Rev. Lett.}\ }\textbf {\bibinfo {volume} {108}},\ \bibinfo {pages}
		{155501} (\bibinfo {year} {2012})}\BibitemShut {NoStop}%
	\bibitem [{\citenamefont {An}\ \emph {et~al.}(2013)\citenamefont {An},
		\citenamefont {Zhang}, \citenamefont {Liu},\ and\ \citenamefont
		{Li}}]{apl102(2013)043113}%
	\BibitemOpen
	\bibfield  {author} {\bibinfo {author} {\bibfnamefont {X.-T.}\ \bibnamefont
			{An}}, \bibinfo {author} {\bibfnamefont {Y.-Y.}\ \bibnamefont {Zhang}},
		\bibinfo {author} {\bibfnamefont {J.-J.}\ \bibnamefont {Liu}}, \ and\
		\bibinfo {author} {\bibfnamefont {S.-S.}\ \bibnamefont {Li}},\ }\href
	{\doibase 10.1063/1.4790147} {\bibfield  {journal} {\bibinfo  {journal}
			{Appl. Phys. Lett.}\ }\textbf {\bibinfo {volume} {102}},\ \bibinfo {pages}
		{043113} (\bibinfo {year} {2013})}\BibitemShut {NoStop}%
	\bibitem [{\citenamefont {Neek-Amal}\ \emph {et~al.}(2013)\citenamefont
		{Neek-Amal}, \citenamefont {Sadeghi}, \citenamefont {Berdiyorov},\ and\
		\citenamefont {Peeters}}]{apl103(2013)261904}%
	\BibitemOpen
	\bibfield  {author} {\bibinfo {author} {\bibfnamefont {M.}~\bibnamefont
			{Neek-Amal}}, \bibinfo {author} {\bibfnamefont {A.}~\bibnamefont {Sadeghi}},
		\bibinfo {author} {\bibfnamefont {G.~R.}\ \bibnamefont {Berdiyorov}}, \ and\
		\bibinfo {author} {\bibfnamefont {F.~M.}\ \bibnamefont {Peeters}},\ }\href
	{\doibase 10.1063/1.4852636} {\bibfield  {journal} {\bibinfo  {journal}
			{Appl. Phys. Lett.}\ }\textbf {\bibinfo {volume} {103}},\ \bibinfo {pages}
		{261904} (\bibinfo {year} {2013})}\BibitemShut {NoStop}%
	\bibitem [{\citenamefont {Li}\ \emph {et~al.}(2013)\citenamefont {Li},
		\citenamefont {Mullen}, \citenamefont {Jin}, \citenamefont {Borysenko},
		\citenamefont {Buongiorno~Nardelli},\ and\ \citenamefont
		{Kim}}]{prb87(2013)115418}%
	\BibitemOpen
	\bibfield  {author} {\bibinfo {author} {\bibfnamefont {X.}~\bibnamefont
			{Li}}, \bibinfo {author} {\bibfnamefont {J.~T.}\ \bibnamefont {Mullen}},
		\bibinfo {author} {\bibfnamefont {Z.}~\bibnamefont {Jin}}, \bibinfo {author}
		{\bibfnamefont {K.~M.}\ \bibnamefont {Borysenko}}, \bibinfo {author}
		{\bibfnamefont {M.}~\bibnamefont {Buongiorno~Nardelli}}, \ and\ \bibinfo
		{author} {\bibfnamefont {K.~W.}\ \bibnamefont {Kim}},\ }\href {\doibase
		10.1103/PhysRevB.87.115418} {\bibfield  {journal} {\bibinfo  {journal} {Phys.
				Rev. B}\ }\textbf {\bibinfo {volume} {87}},\ \bibinfo {pages} {115418}
		(\bibinfo {year} {2013})}\BibitemShut {NoStop}%
	\bibitem [{\citenamefont {Tao}\ \emph {et~al.}(2015)\citenamefont {Tao},
		\citenamefont {Cinquanta}, \citenamefont {Chiappe}, \citenamefont
		{Grazianetti}, \citenamefont {Fanciulli}, \citenamefont {Dubey},
		\citenamefont {Molle},\ and\ \citenamefont {Akinwande}}]{nnano10(2015)227}%
	\BibitemOpen
	\bibfield  {author} {\bibinfo {author} {\bibfnamefont {L.}~\bibnamefont
			{Tao}}, \bibinfo {author} {\bibfnamefont {E.}~\bibnamefont {Cinquanta}},
		\bibinfo {author} {\bibfnamefont {D.}~\bibnamefont {Chiappe}}, \bibinfo
		{author} {\bibfnamefont {C.}~\bibnamefont {Grazianetti}}, \bibinfo {author}
		{\bibfnamefont {M.}~\bibnamefont {Fanciulli}}, \bibinfo {author}
		{\bibfnamefont {M.}~\bibnamefont {Dubey}}, \bibinfo {author} {\bibfnamefont
			{A.}~\bibnamefont {Molle}}, \ and\ \bibinfo {author} {\bibfnamefont
			{D.}~\bibnamefont {Akinwande}},\ }\href
	{https://doi.org/10.1038/nnano.2014.325} {\bibfield  {journal} {\bibinfo
			{journal} {Nat. Nanotechnol.}\ }\textbf {\bibinfo {volume} {10}},\ \bibinfo
		{pages} {227} (\bibinfo {year} {2015})}\BibitemShut {NoStop}%
	\bibitem [{\citenamefont {Le~Lay}(2015)}]{nnano10(2015)202}%
	\BibitemOpen
	\bibfield  {author} {\bibinfo {author} {\bibfnamefont {G.}~\bibnamefont
			{Le~Lay}},\ }\href {https://doi.org/10.1038/nnano.2015.10} {\bibfield
		{journal} {\bibinfo  {journal} {Nat. Nanotechnol.}\ }\textbf {\bibinfo
			{volume} {10}},\ \bibinfo {pages} {202} (\bibinfo {year} {2015})}\BibitemShut
	{NoStop}%
	\bibitem [{\citenamefont {Yeoh}\ \emph {et~al.}(2016)\citenamefont {Yeoh},
		\citenamefont {Ong}, \citenamefont {Ooi}, \citenamefont {Yong},\ and\
		\citenamefont {Lim}}]{sst31(2016)065012}%
	\BibitemOpen
	\bibfield  {author} {\bibinfo {author} {\bibfnamefont {K.~H.}\ \bibnamefont
			{Yeoh}}, \bibinfo {author} {\bibfnamefont {D.~S.}\ \bibnamefont {Ong}},
		\bibinfo {author} {\bibfnamefont {C.~H.~R.}\ \bibnamefont {Ooi}}, \bibinfo
		{author} {\bibfnamefont {T.~K.}\ \bibnamefont {Yong}}, \ and\ \bibinfo
		{author} {\bibfnamefont {S.~K.}\ \bibnamefont {Lim}},\ }\href
	{http://dx.doi.org/10.1088/0268-1242/31/6/065012} {\bibfield  {journal}
		{\bibinfo  {journal} {Semicond. Sci. Technol.}\ }\textbf {\bibinfo {volume}
			{31}},\ \bibinfo {pages} {065012} (\bibinfo {year} {2016})}\BibitemShut
	{NoStop}%
	\bibitem [{\citenamefont {Borowik}, \citenamefont {Thobel},\ and\ \citenamefont
		{Adamowicz}(2016)}]{sst31(2016)115004}%
	\BibitemOpen
	\bibfield  {author} {\bibinfo {author} {\bibfnamefont {P.}~\bibnamefont
			{Borowik}}, \bibinfo {author} {\bibfnamefont {J.-L.}\ \bibnamefont {Thobel}},
		\ and\ \bibinfo {author} {\bibfnamefont {L.}~\bibnamefont {Adamowicz}},\
	}\href {http://dx.doi.org/10.1088/0268-1242/31/11/115004} {\bibfield
		{journal} {\bibinfo  {journal} {Semicond. Sci. Technol.}\ }\textbf {\bibinfo
			{volume} {31}},\ \bibinfo {pages} {115004} (\bibinfo {year}
		{2016})}\BibitemShut {NoStop}%
	\bibitem [{\citenamefont {Gunst}\ \emph {et~al.}(2016)\citenamefont {Gunst},
		\citenamefont {Markussen}, \citenamefont {Stokbro},\ and\ \citenamefont
		{Brandbyge}}]{prb93(2016)035414}%
	\BibitemOpen
	\bibfield  {author} {\bibinfo {author} {\bibfnamefont {T.}~\bibnamefont
			{Gunst}}, \bibinfo {author} {\bibfnamefont {T.}~\bibnamefont {Markussen}},
		\bibinfo {author} {\bibfnamefont {K.}~\bibnamefont {Stokbro}}, \ and\
		\bibinfo {author} {\bibfnamefont {M.}~\bibnamefont {Brandbyge}},\ }\href
	{\doibase 10.1103/PhysRevB.93.035414} {\bibfield  {journal} {\bibinfo
			{journal} {Phys. Rev. B}\ }\textbf {\bibinfo {volume} {93}},\ \bibinfo
		{pages} {035414} (\bibinfo {year} {2016})}\BibitemShut {NoStop}%
	\bibitem [{\citenamefont {Fischetti}\ and\ \citenamefont
		{Vandenberghe}(2016)}]{prb93(2016)155413}%
	\BibitemOpen
	\bibfield  {author} {\bibinfo {author} {\bibfnamefont {M.~V.}\ \bibnamefont
			{Fischetti}}\ and\ \bibinfo {author} {\bibfnamefont {W.~G.}\ \bibnamefont
			{Vandenberghe}},\ }\href {\doibase 10.1103/PhysRevB.93.155413} {\bibfield
		{journal} {\bibinfo  {journal} {Phys. Rev. B}\ }\textbf {\bibinfo {volume}
			{93}},\ \bibinfo {pages} {155413} (\bibinfo {year} {2016})}\BibitemShut
	{NoStop}%
	\bibitem [{\citenamefont {Rengel}\ \emph {et~al.}(2018)\citenamefont {Rengel},
		\citenamefont {Iglesias}, \citenamefont {Hamham},\ and\ \citenamefont
		{Mart\'{i}n}}]{sst33(2018)065011}%
	\BibitemOpen
	\bibfield  {author} {\bibinfo {author} {\bibfnamefont {R.}~\bibnamefont
			{Rengel}}, \bibinfo {author} {\bibfnamefont {J.~M.}\ \bibnamefont
			{Iglesias}}, \bibinfo {author} {\bibfnamefont {E.~M.}\ \bibnamefont
			{Hamham}}, \ and\ \bibinfo {author} {\bibfnamefont {M.~J.}\ \bibnamefont
			{Mart\'{i}n}},\ }\href {http://dx.doi.org/10.1088/1361-6641/aac0a2}
	{\bibfield  {journal} {\bibinfo  {journal} {Semicond. Sci. Technol.}\
		}\textbf {\bibinfo {volume} {33}},\ \bibinfo {pages} {065011} (\bibinfo
		{year} {2018})}\BibitemShut {NoStop}%
	\bibitem [{\citenamefont {Chen}, \citenamefont {Zhong},\ and\ \citenamefont
		{Weinert}(2016)}]{prb94(2016)075409}%
	\BibitemOpen
	\bibfield  {author} {\bibinfo {author} {\bibfnamefont {M.~X.}\ \bibnamefont
			{Chen}}, \bibinfo {author} {\bibfnamefont {Z.}~\bibnamefont {Zhong}}, \ and\
		\bibinfo {author} {\bibfnamefont {M.}~\bibnamefont {Weinert}},\ }\href
	{\doibase 10.1103/PhysRevB.94.075409} {\bibfield  {journal} {\bibinfo
			{journal} {Phys. Rev. B}\ }\textbf {\bibinfo {volume} {94}},\ \bibinfo
		{pages} {075409} (\bibinfo {year} {2016})}\BibitemShut {NoStop}%
	\bibitem [{\citenamefont {Zhao}\ \emph {et~al.}(2016)\citenamefont {Zhao},
		\citenamefont {Liu}, \citenamefont {Yu}, \citenamefont {Quhe}, \citenamefont
		{Zhou}, \citenamefont {Wang}, \citenamefont {Liu}, \citenamefont {Zhong},
		\citenamefont {Han}, \citenamefont {Lu}, \citenamefont {Yao},\ and\
		\citenamefont {Wu}}]{pms83(2016)24}%
	\BibitemOpen
	\bibfield  {author} {\bibinfo {author} {\bibfnamefont {J.}~\bibnamefont
			{Zhao}}, \bibinfo {author} {\bibfnamefont {H.}~\bibnamefont {Liu}}, \bibinfo
		{author} {\bibfnamefont {Z.}~\bibnamefont {Yu}}, \bibinfo {author}
		{\bibfnamefont {R.}~\bibnamefont {Quhe}}, \bibinfo {author} {\bibfnamefont
			{S.}~\bibnamefont {Zhou}}, \bibinfo {author} {\bibfnamefont {Y.}~\bibnamefont
			{Wang}}, \bibinfo {author} {\bibfnamefont {C.~C.}\ \bibnamefont {Liu}},
		\bibinfo {author} {\bibfnamefont {H.}~\bibnamefont {Zhong}}, \bibinfo
		{author} {\bibfnamefont {N.}~\bibnamefont {Han}}, \bibinfo {author}
		{\bibfnamefont {J.}~\bibnamefont {Lu}}, \bibinfo {author} {\bibfnamefont
			{Y.}~\bibnamefont {Yao}}, \ and\ \bibinfo {author} {\bibfnamefont
			{K.}~\bibnamefont {Wu}},\ }\href
	{http://www.sciencedirect.com/science/article/pii/S0079642516300068}
	{\bibfield  {journal} {\bibinfo  {journal} {Prog. Mater Sci.}\ }\textbf
		{\bibinfo {volume} {83}},\ \bibinfo {pages} {24} (\bibinfo {year}
		{2016})}\BibitemShut {NoStop}%
	\bibitem [{\citenamefont {Gaddemane}\ \emph {et~al.}(2018)\citenamefont
		{Gaddemane}, \citenamefont {Vandenberghe}, \citenamefont {Van~de Put},
		\citenamefont {Chen},\ and\ \citenamefont {Fischetti}}]{jap124(2018)044306}%
	\BibitemOpen
	\bibfield  {author} {\bibinfo {author} {\bibfnamefont {G.}~\bibnamefont
			{Gaddemane}}, \bibinfo {author} {\bibfnamefont {W.~G.}\ \bibnamefont
			{Vandenberghe}}, \bibinfo {author} {\bibfnamefont {M.~L.}\ \bibnamefont
			{Van~de Put}}, \bibinfo {author} {\bibfnamefont {E.}~\bibnamefont {Chen}}, \
		and\ \bibinfo {author} {\bibfnamefont {M.~V.}\ \bibnamefont {Fischetti}},\
	}\href {\doibase 10.1063/1.5037581} {\bibfield  {journal} {\bibinfo
			{journal} {J. Appl. Phys.}\ }\textbf {\bibinfo {volume} {124}},\ \bibinfo
		{pages} {044306} (\bibinfo {year} {2018})}\BibitemShut {NoStop}%
	\bibitem [{\citenamefont {\"{O}zdemir}\ \emph {et~al.}(2020)\citenamefont
		{\"{O}zdemir}, \citenamefont {Cekil}, \citenamefont {Atasever}, \citenamefont
		{\"{O}zdemir}, \citenamefont {Yarar},\ and\ \citenamefont
		{\"{O}zdemir}}]{jms1199(2020)126878}%
	\BibitemOpen
	\bibfield  {author} {\bibinfo {author} {\bibfnamefont {M.~D.}\ \bibnamefont
			{\"{O}zdemir}}, \bibinfo {author} {\bibfnamefont {H.~C.}\ \bibnamefont
			{Cekil}}, \bibinfo {author} {\bibfnamefont {O.}~\bibnamefont {Atasever}},
		\bibinfo {author} {\bibfnamefont {B.}~\bibnamefont {\"{O}zdemir}}, \bibinfo
		{author} {\bibfnamefont {Z.}~\bibnamefont {Yarar}}, \ and\ \bibinfo {author}
		{\bibfnamefont {M.}~\bibnamefont {\"{O}zdemir}},\ }\href
	{http://www.sciencedirect.com/science/article/pii/S002228601930969X}
	{\bibfield  {journal} {\bibinfo  {journal} {J. Mol. Struct.}\ }\textbf
		{\bibinfo {volume} {1199}},\ \bibinfo {pages} {126878} (\bibinfo {year}
		{2020})}\BibitemShut {NoStop}%
	\bibitem [{\citenamefont {Muoi}\ \emph {et~al.}(2020)\citenamefont {Muoi},
		\citenamefont {Hieu}, \citenamefont {Nguyen}, \citenamefont {Hoi},
		\citenamefont {Nguyen}, \citenamefont {Hien}, \citenamefont {Poklonski},
		\citenamefont {Kubakaddi},\ and\ \citenamefont {Phuc}}]{prb101(2020)205408}%
	\BibitemOpen
	\bibfield  {author} {\bibinfo {author} {\bibfnamefont {D.}~\bibnamefont
			{Muoi}}, \bibinfo {author} {\bibfnamefont {N.~N.}\ \bibnamefont {Hieu}},
		\bibinfo {author} {\bibfnamefont {C.~V.}\ \bibnamefont {Nguyen}}, \bibinfo
		{author} {\bibfnamefont {B.~D.}\ \bibnamefont {Hoi}}, \bibinfo {author}
		{\bibfnamefont {H.~V.}\ \bibnamefont {Nguyen}}, \bibinfo {author}
		{\bibfnamefont {N.~D.}\ \bibnamefont {Hien}}, \bibinfo {author}
		{\bibfnamefont {N.~A.}\ \bibnamefont {Poklonski}}, \bibinfo {author}
		{\bibfnamefont {S.~S.}\ \bibnamefont {Kubakaddi}}, \ and\ \bibinfo {author}
		{\bibfnamefont {H.~V.}\ \bibnamefont {Phuc}},\ }\href {\doibase
		10.1103/PhysRevB.101.205408} {\bibfield  {journal} {\bibinfo  {journal}
			{Phys. Rev. B}\ }\textbf {\bibinfo {volume} {101}},\ \bibinfo {pages}
		{205408} (\bibinfo {year} {2020})}\BibitemShut {NoStop}%
	\bibitem [{\citenamefont {Hwang}\ and\ \citenamefont
		{Das~Sarma}(2008)}]{prb77(2008)115449}%
	\BibitemOpen
	\bibfield  {author} {\bibinfo {author} {\bibfnamefont {E.~H.}\ \bibnamefont
			{Hwang}}\ and\ \bibinfo {author} {\bibfnamefont {S.}~\bibnamefont
			{Das~Sarma}},\ }\href {\doibase 10.1103/PhysRevB.77.115449} {\bibfield
		{journal} {\bibinfo  {journal} {Phys. Rev. B}\ }\textbf {\bibinfo {volume}
			{77}},\ \bibinfo {pages} {115449} (\bibinfo {year} {2008})}\BibitemShut
	{NoStop}%
	\bibitem [{\citenamefont {Kubakaddi}(2009)}]{prb79(2009)075417}%
	\BibitemOpen
	\bibfield  {author} {\bibinfo {author} {\bibfnamefont {S.~S.}\ \bibnamefont
			{Kubakaddi}},\ }\href {\doibase 10.1103/PhysRevB.79.075417} {\bibfield
		{journal} {\bibinfo  {journal} {Phys. Rev. B}\ }\textbf {\bibinfo {volume}
			{79}},\ \bibinfo {pages} {075417} (\bibinfo {year} {2009})}\BibitemShut
	{NoStop}%
	\bibitem [{\citenamefont {Tse}\ and\ \citenamefont
		{Das~Sarma}(2009)}]{prb79(2009)235406}%
	\BibitemOpen
	\bibfield  {author} {\bibinfo {author} {\bibfnamefont {W.-K.}\ \bibnamefont
			{Tse}}\ and\ \bibinfo {author} {\bibfnamefont {S.}~\bibnamefont
			{Das~Sarma}},\ }\href {\doibase 10.1103/PhysRevB.79.235406} {\bibfield
		{journal} {\bibinfo  {journal} {Phys. Rev. B}\ }\textbf {\bibinfo {volume}
			{79}},\ \bibinfo {pages} {235406} (\bibinfo {year} {2009})}\BibitemShut
	{NoStop}%
	\bibitem [{\citenamefont {Viljas}\ and\ \citenamefont
		{Heikkil\"a}(2010)}]{prb81(2010)245404}%
	\BibitemOpen
	\bibfield  {author} {\bibinfo {author} {\bibfnamefont {J.~K.}\ \bibnamefont
			{Viljas}}\ and\ \bibinfo {author} {\bibfnamefont {T.~T.}\ \bibnamefont
			{Heikkil\"a}},\ }\href {\doibase 10.1103/PhysRevB.81.245404} {\bibfield
		{journal} {\bibinfo  {journal} {Phys. Rev. B}\ }\textbf {\bibinfo {volume}
			{81}},\ \bibinfo {pages} {245404} (\bibinfo {year} {2010})}\BibitemShut
	{NoStop}%
	\bibitem [{\citenamefont {Efetov}\ and\ \citenamefont
		{Kim}(2010)}]{prl105(2010)256805}%
	\BibitemOpen
	\bibfield  {author} {\bibinfo {author} {\bibfnamefont {D.~K.}\ \bibnamefont
			{Efetov}}\ and\ \bibinfo {author} {\bibfnamefont {P.}~\bibnamefont {Kim}},\
	}\href {\doibase 10.1103/PhysRevLett.105.256805} {\bibfield  {journal}
		{\bibinfo  {journal} {Phys. Rev. Lett.}\ }\textbf {\bibinfo {volume} {105}},\
		\bibinfo {pages} {256805} (\bibinfo {year} {2010})}\BibitemShut {NoStop}%
	\bibitem [{\citenamefont {Betz}\ \emph {et~al.}(2012)\citenamefont {Betz},
		\citenamefont {Vialla}, \citenamefont {Brunel}, \citenamefont {Voisin},
		\citenamefont {Picher}, \citenamefont {Cavanna}, \citenamefont {Madouri},
		\citenamefont {F\`eve}, \citenamefont {Berroir}, \citenamefont
		{Pla\ifmmode~\mbox{\c{c}}\else \c{c}\fi{}ais},\ and\ \citenamefont
		{Pallecchi}}]{prl109(2012)056805}%
	\BibitemOpen
	\bibfield  {author} {\bibinfo {author} {\bibfnamefont {A.~C.}\ \bibnamefont
			{Betz}}, \bibinfo {author} {\bibfnamefont {F.}~\bibnamefont {Vialla}},
		\bibinfo {author} {\bibfnamefont {D.}~\bibnamefont {Brunel}}, \bibinfo
		{author} {\bibfnamefont {C.}~\bibnamefont {Voisin}}, \bibinfo {author}
		{\bibfnamefont {M.}~\bibnamefont {Picher}}, \bibinfo {author} {\bibfnamefont
			{A.}~\bibnamefont {Cavanna}}, \bibinfo {author} {\bibfnamefont
			{A.}~\bibnamefont {Madouri}}, \bibinfo {author} {\bibfnamefont
			{G.}~\bibnamefont {F\`eve}}, \bibinfo {author} {\bibfnamefont {J.-M.}\
			\bibnamefont {Berroir}}, \bibinfo {author} {\bibfnamefont {B.}~\bibnamefont
			{Pla\ifmmode~\mbox{\c{c}}\else \c{c}\fi{}ais}}, \ and\ \bibinfo {author}
		{\bibfnamefont {E.}~\bibnamefont {Pallecchi}},\ }\href {\doibase
		10.1103/PhysRevLett.109.056805} {\bibfield  {journal} {\bibinfo  {journal}
			{Phys. Rev. Lett.}\ }\textbf {\bibinfo {volume} {109}},\ \bibinfo {pages}
		{056805} (\bibinfo {year} {2012})}\BibitemShut {NoStop}%
	\bibitem [{\citenamefont {Baker}\ \emph {et~al.}(2012)\citenamefont {Baker},
		\citenamefont {Alexander-Webber}, \citenamefont {Altebaeumer},\ and\
		\citenamefont {Nicholas}}]{prb85(2012)115403}%
	\BibitemOpen
	\bibfield  {author} {\bibinfo {author} {\bibfnamefont {A.~M.~R.}\
			\bibnamefont {Baker}}, \bibinfo {author} {\bibfnamefont {J.~A.}\ \bibnamefont
			{Alexander-Webber}}, \bibinfo {author} {\bibfnamefont {T.}~\bibnamefont
			{Altebaeumer}}, \ and\ \bibinfo {author} {\bibfnamefont {R.~J.}\ \bibnamefont
			{Nicholas}},\ }\href {\doibase 10.1103/PhysRevB.85.115403} {\bibfield
		{journal} {\bibinfo  {journal} {Phys. Rev. B}\ }\textbf {\bibinfo {volume}
			{85}},\ \bibinfo {pages} {115403} (\bibinfo {year} {2012})}\BibitemShut
	{NoStop}%
	\bibitem [{\citenamefont {Low}\ \emph {et~al.}(2012)\citenamefont {Low},
		\citenamefont {Perebeinos}, \citenamefont {Kim}, \citenamefont {Freitag},\
		and\ \citenamefont {Avouris}}]{prb86(2012)045413}%
	\BibitemOpen
	\bibfield  {author} {\bibinfo {author} {\bibfnamefont {T.}~\bibnamefont
			{Low}}, \bibinfo {author} {\bibfnamefont {V.}~\bibnamefont {Perebeinos}},
		\bibinfo {author} {\bibfnamefont {R.}~\bibnamefont {Kim}}, \bibinfo {author}
		{\bibfnamefont {M.}~\bibnamefont {Freitag}}, \ and\ \bibinfo {author}
		{\bibfnamefont {P.}~\bibnamefont {Avouris}},\ }\href {\doibase
		10.1103/PhysRevB.86.045413} {\bibfield  {journal} {\bibinfo  {journal} {Phys.
				Rev. B}\ }\textbf {\bibinfo {volume} {86}},\ \bibinfo {pages} {045413}
		(\bibinfo {year} {2012})}\BibitemShut {NoStop}%
	\bibitem [{\citenamefont {Katti}\ and\ \citenamefont
		{Kubakaddi}(2013)}]{jap113(2013)063705}%
	\BibitemOpen
	\bibfield  {author} {\bibinfo {author} {\bibfnamefont {V.~S.}\ \bibnamefont
			{Katti}}\ and\ \bibinfo {author} {\bibfnamefont {S.~S.}\ \bibnamefont
			{Kubakaddi}},\ }\href {\doibase 10.1063/1.4790309} {\bibfield  {journal}
		{\bibinfo  {journal} {J. Appl. Phys.}\ }\textbf {\bibinfo {volume} {113}},\
		\bibinfo {pages} {063705} (\bibinfo {year} {2013})}\BibitemShut {NoStop}%
	\bibitem [{\citenamefont {Kubakaddi}\ and\ \citenamefont
		{Biswas}(2018)}]{jpcm30(2018)265303}%
	\BibitemOpen
	\bibfield  {author} {\bibinfo {author} {\bibfnamefont {S.~S.}\ \bibnamefont
			{Kubakaddi}}\ and\ \bibinfo {author} {\bibfnamefont {T.}~\bibnamefont
			{Biswas}},\ }\href {http://dx.doi.org/10.1088/1361-648X/aac661} {\bibfield
		{journal} {\bibinfo  {journal} {J. Phys.: Condens. Matter}\ }\textbf
		{\bibinfo {volume} {30}},\ \bibinfo {pages} {265303} (\bibinfo {year}
		{2018})}\BibitemShut {NoStop}%
	\bibitem [{\citenamefont {Perebeinos}\ and\ \citenamefont
		{Avouris}(2010)}]{prb81(2010)195442}%
	\BibitemOpen
	\bibfield  {author} {\bibinfo {author} {\bibfnamefont {V.}~\bibnamefont
			{Perebeinos}}\ and\ \bibinfo {author} {\bibfnamefont {P.}~\bibnamefont
			{Avouris}},\ }\href {\doibase 10.1103/PhysRevB.81.195442} {\bibfield
		{journal} {\bibinfo  {journal} {Phys. Rev. B}\ }\textbf {\bibinfo {volume}
			{81}},\ \bibinfo {pages} {195442} (\bibinfo {year} {2010})}\BibitemShut
	{NoStop}%
	\bibitem [{\citenamefont {DaSilva}\ \emph {et~al.}(2010)\citenamefont
		{DaSilva}, \citenamefont {Zou}, \citenamefont {Jain},\ and\ \citenamefont
		{Zhu}}]{prl104(2010)236601}%
	\BibitemOpen
	\bibfield  {author} {\bibinfo {author} {\bibfnamefont {A.~M.}\ \bibnamefont
			{DaSilva}}, \bibinfo {author} {\bibfnamefont {K.}~\bibnamefont {Zou}},
		\bibinfo {author} {\bibfnamefont {J.~K.}\ \bibnamefont {Jain}}, \ and\
		\bibinfo {author} {\bibfnamefont {J.}~\bibnamefont {Zhu}},\ }\href {\doibase
		10.1103/PhysRevLett.104.236601} {\bibfield  {journal} {\bibinfo  {journal}
			{Phys. Rev. Lett.}\ }\textbf {\bibinfo {volume} {104}},\ \bibinfo {pages}
		{236601} (\bibinfo {year} {2010})}\BibitemShut {NoStop}%
	\bibitem [{\citenamefont {Konar}, \citenamefont {Fang},\ and\ \citenamefont
		{Jena}(2010)}]{prb82(2010)115452}%
	\BibitemOpen
	\bibfield  {author} {\bibinfo {author} {\bibfnamefont {A.}~\bibnamefont
			{Konar}}, \bibinfo {author} {\bibfnamefont {T.}~\bibnamefont {Fang}}, \ and\
		\bibinfo {author} {\bibfnamefont {D.}~\bibnamefont {Jena}},\ }\href {\doibase
		10.1103/PhysRevB.82.115452} {\bibfield  {journal} {\bibinfo  {journal} {Phys.
				Rev. B}\ }\textbf {\bibinfo {volume} {82}},\ \bibinfo {pages} {115452}
		(\bibinfo {year} {2010})}\BibitemShut {NoStop}%
	\bibitem [{\citenamefont {Fischetti}, \citenamefont {Neumayer},\ and\
		\citenamefont {Cartier}(2001)}]{jap90(2001)4587}%
	\BibitemOpen
	\bibfield  {author} {\bibinfo {author} {\bibfnamefont {M.~V.}\ \bibnamefont
			{Fischetti}}, \bibinfo {author} {\bibfnamefont {D.~A.}\ \bibnamefont
			{Neumayer}}, \ and\ \bibinfo {author} {\bibfnamefont {E.~A.}\ \bibnamefont
			{Cartier}},\ }\href {\doibase 10.1063/1.1405826} {\bibfield  {journal}
		{\bibinfo  {journal} {J. Appl. Phys.}\ }\textbf {\bibinfo {volume} {90}},\
		\bibinfo {pages} {4587} (\bibinfo {year} {2001})}\BibitemShut {NoStop}%
	\bibitem [{\citenamefont {Hamham}\ \emph {et~al.}(2018)\citenamefont {Hamham},
		\citenamefont {Iglesias}, \citenamefont {Pascual}, \citenamefont
		{Mart\'{i}n},\ and\ \citenamefont {Rengel}}]{jpdap51(2018)415102}%
	\BibitemOpen
	\bibfield  {author} {\bibinfo {author} {\bibfnamefont {E.~M.}\ \bibnamefont
			{Hamham}}, \bibinfo {author} {\bibfnamefont {J.~M.}\ \bibnamefont
			{Iglesias}}, \bibinfo {author} {\bibfnamefont {E.}~\bibnamefont {Pascual}},
		\bibinfo {author} {\bibfnamefont {M.~J.}\ \bibnamefont {Mart\'{i}n}}, \ and\
		\bibinfo {author} {\bibfnamefont {R.}~\bibnamefont {Rengel}},\ }\href
	{http://dx.doi.org/10.1088/1361-6463/aad94c} {\bibfield  {journal} {\bibinfo
			{journal} {J. Phys. D: Appl. Phys.}\ }\textbf {\bibinfo {volume} {51}},\
		\bibinfo {pages} {415102} (\bibinfo {year} {2018})}\BibitemShut {NoStop}%
	\bibitem [{\citenamefont {Baker}\ \emph {et~al.}(2013)\citenamefont {Baker},
		\citenamefont {Alexander-Webber}, \citenamefont {Altebaeumer}, \citenamefont
		{McMullan}, \citenamefont {Janssen}, \citenamefont {Tzalenchuk},
		\citenamefont {Lara-Avila}, \citenamefont {Kubatkin}, \citenamefont
		{Yakimova}, \citenamefont {Lin}, \citenamefont {Li},\ and\ \citenamefont
		{Nicholas}}]{prb87(2013)045414}%
	\BibitemOpen
	\bibfield  {author} {\bibinfo {author} {\bibfnamefont {A.~M.~R.}\
			\bibnamefont {Baker}}, \bibinfo {author} {\bibfnamefont {J.~A.}\ \bibnamefont
			{Alexander-Webber}}, \bibinfo {author} {\bibfnamefont {T.}~\bibnamefont
			{Altebaeumer}}, \bibinfo {author} {\bibfnamefont {S.~D.}\ \bibnamefont
			{McMullan}}, \bibinfo {author} {\bibfnamefont {T.~J. B.~M.}\ \bibnamefont
			{Janssen}}, \bibinfo {author} {\bibfnamefont {A.}~\bibnamefont {Tzalenchuk}},
		\bibinfo {author} {\bibfnamefont {S.}~\bibnamefont {Lara-Avila}}, \bibinfo
		{author} {\bibfnamefont {S.}~\bibnamefont {Kubatkin}}, \bibinfo {author}
		{\bibfnamefont {R.}~\bibnamefont {Yakimova}}, \bibinfo {author}
		{\bibfnamefont {C.-T.}\ \bibnamefont {Lin}}, \bibinfo {author} {\bibfnamefont
			{L.-J.}\ \bibnamefont {Li}}, \ and\ \bibinfo {author} {\bibfnamefont {R.~J.}\
			\bibnamefont {Nicholas}},\ }\href {\doibase 10.1103/PhysRevB.87.045414}
	{\bibfield  {journal} {\bibinfo  {journal} {Phys. Rev. B}\ }\textbf {\bibinfo
			{volume} {87}},\ \bibinfo {pages} {045414} (\bibinfo {year}
		{2013})}\BibitemShut {NoStop}%
	\bibitem [{\citenamefont {Huang}\ \emph {et~al.}(2015)\citenamefont {Huang},
		\citenamefont {Alexander-Webber}, \citenamefont {Janssen}, \citenamefont
		{Tzalenchuk}, \citenamefont {Yager}, \citenamefont {Lara-Avila},
		\citenamefont {Kubatkin}, \citenamefont {Myers-Ward}, \citenamefont
		{Wheeler}, \citenamefont {Gaskill},\ and\ \citenamefont
		{Nicholas}}]{jpcm27(2015)164202}%
	\BibitemOpen
	\bibfield  {author} {\bibinfo {author} {\bibfnamefont {J.}~\bibnamefont
			{Huang}}, \bibinfo {author} {\bibfnamefont {J.~A.}\ \bibnamefont
			{Alexander-Webber}}, \bibinfo {author} {\bibfnamefont {T.~J. B.~M.}\
			\bibnamefont {Janssen}}, \bibinfo {author} {\bibfnamefont {A.}~\bibnamefont
			{Tzalenchuk}}, \bibinfo {author} {\bibfnamefont {T.}~\bibnamefont {Yager}},
		\bibinfo {author} {\bibfnamefont {S.}~\bibnamefont {Lara-Avila}}, \bibinfo
		{author} {\bibfnamefont {S.}~\bibnamefont {Kubatkin}}, \bibinfo {author}
		{\bibfnamefont {R.~L.}\ \bibnamefont {Myers-Ward}}, \bibinfo {author}
		{\bibfnamefont {V.~D.}\ \bibnamefont {Wheeler}}, \bibinfo {author}
		{\bibfnamefont {D.~K.}\ \bibnamefont {Gaskill}}, \ and\ \bibinfo {author}
		{\bibfnamefont {R.~J.}\ \bibnamefont {Nicholas}},\ }\href
	{http://dx.doi.org/10.1088/0953-8984/27/16/164202} {\bibfield  {journal}
		{\bibinfo  {journal} {J. Phys.: Condens. Matter}\ }\textbf {\bibinfo {volume}
			{27}},\ \bibinfo {pages} {164202} (\bibinfo {year} {2015})}\BibitemShut
	{NoStop}%
	\bibitem [{\citenamefont {Kaasbjerg}, \citenamefont {Bhargavi},\ and\
		\citenamefont {Kubakaddi}(2014)}]{prb90(2014)165436}%
	\BibitemOpen
	\bibfield  {author} {\bibinfo {author} {\bibfnamefont {K.}~\bibnamefont
			{Kaasbjerg}}, \bibinfo {author} {\bibfnamefont {K.~S.}\ \bibnamefont
			{Bhargavi}}, \ and\ \bibinfo {author} {\bibfnamefont {S.~S.}\ \bibnamefont
			{Kubakaddi}},\ }\href {\doibase 10.1103/PhysRevB.90.165436} {\bibfield
		{journal} {\bibinfo  {journal} {Phys. Rev. B}\ }\textbf {\bibinfo {volume}
			{90}},\ \bibinfo {pages} {165436} (\bibinfo {year} {2014})}\BibitemShut
	{NoStop}%
	\bibitem [{\citenamefont {Shishir}\ and\ \citenamefont
		{Ferry}(2009)}]{jpcm21(2009)232204}%
	\BibitemOpen
	\bibfield  {author} {\bibinfo {author} {\bibfnamefont {R.~S.}\ \bibnamefont
			{Shishir}}\ and\ \bibinfo {author} {\bibfnamefont {D.~K.}\ \bibnamefont
			{Ferry}},\ }\href {http://dx.doi.org/10.1088/0953-8984/21/23/232204}
	{\bibfield  {journal} {\bibinfo  {journal} {J. Phys.: Condens. Matter}\
		}\textbf {\bibinfo {volume} {21}},\ \bibinfo {pages} {232204} (\bibinfo
		{year} {2009})}\BibitemShut {NoStop}%
	\bibitem [{\citenamefont {Rengel}()}]{Rengel}%
	\BibitemOpen
	\bibfield  {author} {\bibinfo {author} {\bibfnamefont {R.}~\bibnamefont
			{Rengel}},\ }\href@noop {} {}\bibinfo {note} {(private
		communication)}\BibitemShut {NoStop}%
	\bibitem [{\citenamefont {Fischetti}()}]{Fischetti}%
	\BibitemOpen
	\bibfield  {author} {\bibinfo {author} {\bibfnamefont {M.~V.}\ \bibnamefont
			{Fischetti}},\ }\href@noop {} {}\bibinfo {note} {(private
		communication)}\BibitemShut {NoStop}%
\end{thebibliography}
\providecommand{\noopsort}[1]{}\providecommand{\singleletter}[1]{#1}%

\end{document}